\documentclass[12pt,leqno]{article}

\usepackage{setspace} 
\usepackage[vmargin = 1.5in, hmargin =1.5in]{geometry}

\usepackage{sectsty,titlecaps,secdot} 
\sectionfont{\centering \fontsize{12}{15}\selectfont \MakeUppercase} 
\subsectionfont{\centering \fontsize{10}{15}\selectfont \MakeUppercase}
\usepackage{amsmath,amssymb,amsthm,amsfonts,mathtools,stmaryrd,xfrac,dsfont,bm,array,multirow,physics, accents,makecell} 
\newcolumntype{M}[1]{>{\centering\arraybackslash}m{#1}}
\usepackage{natbib} 
\usepackage{accents}
\newcommand{\ubar}[1]{\underaccent{\bar}{#1}}
\usepackage{needspace}
\def\citeapos#1{\citeauthor{#1}'s (\citeyear{#1})}
\usepackage{graphicx} 
\usepackage[utf8]{inputenc} 
\usepackage{babel} 
\usepackage{float,framed,tikz,tkz-euclide,subcaption} 
\usepackage[shortlabels]{enumitem} 
\usepackage[titletoc]{appendix} 
\usepackage{verbatim} 
\usetikzlibrary{matrix}
\usetikzlibrary{arrows}
\usetikzlibrary{patterns}
\usetikzlibrary{calc}
\usetikzlibrary{shapes.misc}
\usetikzlibrary{positioning,decorations.pathreplacing,shapes}
\usetikzlibrary{decorations.pathreplacing}
\usepackage{pgfplotstable}
\usepackage{pgfplots}
\pgfplotsset{compat=1.12}
\usepgfplotslibrary{fillbetween}
\usepackage[hidelinks]{hyperref} 
\definecolor{Blue}{RGB}{86,180,233}
\hypersetup{colorlinks,linkcolor={Blue},citecolor={Blue}}  

\usepackage{pdfpages} 
\usepackage{pgfplots}
\usepgfplotslibrary{groupplots,colorbrewer}
\pgfplotsset{compat=newest}
\pgfplotsset{cycle list/Set1}
\usepackage{tikz}
\usetikzlibrary{matrix,calc,shapes,arrows.meta,positioning}
\tikzset{
    vertex/.style = {shape=circle,draw, minimum size = 1.8em, inner sep = 0pt},
    edge/.style = {->,> = latex}
}

\definecolor{Blue}{RGB}{86,180,233}
\definecolor{Orange}{RGB}{230,159,0}
\definecolor{Green}{RGB}{0,158,115}

\usepackage[capitalize,noabbrev]{cleveref}

\newtheoremstyle{break}
{}
{}
{\itshape}
{}
{\bfseries}
{}
{\newline}
{}

\theoremstyle{break}
\newtheorem{thm}{Theorem}
\newtheorem*{theorem*}{Theorem}
\newtheorem*{cor*}{Corollary}
\newtheorem{cor}{Corollary}

\newtheorem{lem}{Lemma}

\crefname{prop}{Proposition}{Propositions}
\crefname{thm}{Theorem}{Theorems}
\crefname{lem}{Lemma}{Lemmas}
\crefname{blem}{lem}{Lemmas}

\theoremstyle{definition}

\newtheorem{rem}{Remark}
\newtheorem*{rem*}{Remark}
\newtheorem*{claim*}{Claim}

\newtheorem{as}{Assumption}
\newtheorem{cond}{Condition}

\def\a{\alpha}

\def\g{\gamma}

\def\e{\varepsilon}
\def\z{\zeta}

\def\th{\theta}

\def\D{\Delta}
\def\Th{\Theta}

\def\R{\mathbf{R}}

\def\HH{\mathcal{H}}

\def\P{\mathbf{P}}

\def\ol{\bar}
\newcommand{\ul}[1]{\underaccent{\bar}{#1}}

\DeclareMathOperator{\E}{\mathbb{E}}

\DeclareMathOperator{\dom}{dom} 

\DeclareMathOperator{\proj}{proj}

\newcommand{\Paren}[1]{\left( #1 \right)}

\newcommand{\Brac}[1]{\left[ #1 \right]}

\newcommand{\Set}[1]{\left\{ #1 \right\}}

\newcommand{\de}{\mathop{}\!\mathrm{d}}

\usepackage{datetime}
\newdateformat{specialdate}{\THEDAY~\monthname[\THEMONTH] \THEYEAR}

\begin{document}

\title{Optimal Auction Design with Contingent Payments and Costly Verification\thanks{Ball: Department of Economics, MIT (ianball@mit.edu); 
Pekkarinen: Department of Economics, University of Vaasa (teemu.pekkarinen@uwasa.fi). This paper was previously circulated under the titles ``Optimal Auction Design with Flexible Royalty Payments'' and ``Optimal Auction Design with Contingent Payments.'' For helpful comments, we are grateful to Dirk Bergemann, Daniel Clark, Laura Doval, Marina Halac, Daniel Hauser, Topi Hokkanen, Johannes Hörner, Deniz Kattwinkel, Jan Knoepfle, Bart Lipman, Laurenz Marstaller, Stephen Morris, Petteri Palonen, Juuso Toikka, Hannu Vartiainen, Juuso Välimäki, Stephan Waizmann, Alex Winter, Alex Wolitzky, and members of the microeconomic theory groups at the Helsinki GSE, MIT, Yale, TSE, and Bonn. Part of this paper was written while Ian Ball was visiting Northwestern University and Teemu Pekkarinen was visiting MIT.}}
\author{Ian Ball \and Teemu Pekkarinen}
\date{16 April 2026}

\maketitle

\begin{abstract}
We study the design of an auction for an income-generating asset such as an intellectual property license. Each bidder has a signal about his future income from acquiring the asset. After the asset is allocated, the winner's income from the asset is realized privately. The principal can audit the winner, at a cost, and then charge a payment contingent on the winner's realized income. We solve for an auction that maximizes the principal's revenue, net of auditing costs. The winning bidder is charged linear royalties up to a cap, beyond which there is no auditing. A higher bidder pays more in cash upfront and faces a lower royalty cap. (JEL D44, D82, D86) 
\end{abstract}



\newpage

\section{Introduction}

In 2017, Cardiva Medical Inc.\ entered into an asset purchase agreement with Interventional Therapies, LLC, a developer of medical devices.\footnote{The full asset purchase agreement is available in the SEC archives at \href{https://www.sec.gov/Archives/edgar/data/1520325/000119312521001047/d833938dex21.htm}{\color{Blue}{https://www.sec.gov/Archives/edgar/data/1520325/000119312521001047/d833938dex21.htm}}.}  Cardiva agreed to pay Interventional Therapies \$100{,}000 at closing and a 5\% royalty on net sales of a specified list of products, subject to a royalty cap of \$7.5 million. To determine the royalties owed, the agreement requires the buyer to submit quarterly sales reports. To verify these reports, the seller may audit the buyer's records using an independent certified public accountant. Any underpayment revealed by an audit must be paid, and any overpayment refunded.

Royalty agreements are standard in various forms of intellectual property licensing such as patents, copyrights, and franchises.\footnote{See \cite{Contreras2022} for a textbook treatment of intellectual-property licensing.}  Under a royalty agreement, the amount that the buyer ultimately pays is contingent on subsequent sales. More generally, contingent payments are common whenever an asset generates verifiable cash flows. When there are multiple potential buyers, contingent-payment auctions are often used, as in government auctions for casino licenses, oil leases, and procurement contracts \citep{skrzypacz2013auctions}.

The theoretical literature on contingent-payment auctions, following \cite{demarzo2005bidding}, assumes that the income generated by the asset is publicly observed. But in many applications, such as in the royalty agreement described above, cash flows are privately observed and can only be verified at a cost. Indeed, an important consideration in patent licensing is that ``a lump-sum royalty removes the administrative burden and costs of monitoring the actual use of the licensed technology because the royalty payment is independent of the licensee’s \emph{actual} sales'' \citep[p.~903]{Sidak2016}. 

In this paper, we study the design of an auction for an asset that generates income, which can be verified at a cost. For example, the asset could specify the right to use some intellectual property (such as a patent, trademark, copyright, or franchise). Or the asset could specify the right to operate in a particular industry (such as a government license for a casino or a marijuana dispensary). Procurement contracts can also be covered by our analysis after an appropriate sign change; costs, rather than income, are subsequently realized,  so contingent payments are interpreted as cost-sharing rather than royalties. 

We consider the following dynamic mechanism design setting. Each agent observes an independent, private signal about his future income from acquiring the asset. As in a standard auction, the principal solicits bids from the agents. Based on these bids, the principal charges upfront transfers and allocates the asset to at most one agent. The next phase is specific to our setting. If the asset is allocated, the winner's income from the asset is realized. The winner privately observes this income and makes a report to the principal. Based on this report, the principal charges the winner a payment. Finally, the principal can conduct a costly audit that reveals the winner's realized income. If the principal conducts the audit, she can then charge the winner a payment that depends on the winner's realized income. The slope of this payment, as a function of income, is bounded, as specified in the model.

We solve for a dynamic mechanism that maximizes the principal's revenue, net of auditing costs. To build intuition for the solution, we begin with two benchmarks. 

If the principal cannot charge payments that are contingent on realized income, then the model reduces to the standard cash-payment auction setting of \cite{myerson1981optimal}. At the other extreme, suppose that the income generated by the asset is public and that the principal can charge \emph{unrestricted} payments contingent on this realized income. In this case, the principal can fully extract the surplus using the following modified first-price auction. The allocation and upfront transfers are the same as in a standard first-price auction, but the winner is charged a penalty equal to the difference between his realized income from the asset and his bid. In this auction, each agent receives zero utility, no matter how he bids. In particular, it is an equilibrium for each agent to bid his expected income from the asset. In this equilibrium, the asset is allocated efficiently, so the principal extracts the full surplus. 

Our main model interpolates between these benchmarks. In \cref{MAINTHEOREM}, we identify an optimal mechanism. To solve for this mechanism, we find an expression for the virtual value in our setting. This virtual value is weakly higher than Myerson's virtual value (for cash auctions) because contingent payments reduce information rents. In our optimal mechanism, the asset is allocated to the agent with the highest virtual value, provided that this virtual value is positive.  If the asset is allocated, the principal charges the winner a royalty based on his reported income. The royalty is linear up to a royalty cap (which depends on the winner's bid). The winner is audited if and only if he claims that the royalty he owes is strictly less than the cap. If auditing reveals that the agent under-reported his income, then he is charged the underpaid royalties as a penalty. The increasing segment of the royalty schedule reduces information rents. Over this segment,  costly auditing is necessary to motivate the winner to report his income truthfully. The flat segment beyond the royalty cap saves on auditing costs. Thus, our analysis helps to rationalize the use of royalty caps in practice, as in the agreement between Cardiva and Interventional Therapies.
 
With a single agent, the optimal auction in \cref{MAINTHEOREM} can be implemented by posting a menu of contracts, each with a different royalty cap and upfront price. The agent chooses from this menu based upon his private signal about future income. 

\cref{VV:COMP} analyzes comparative statics. Our optimal auction interpolates between the Myerson optimal cash auction  and full-surplus extraction. If auditing is cheaper or if contingent payments are allowed to be steeper, then the virtual value increases, and royalty caps increase.  For example, if technological advances reduce auditing costs, our model predicts that royalty caps will increase. 



\subsection*{Related Literature}

We depart from previous work on contingent-payment auctions by modeling the winner's realized income as private information that can only be verified at a cost.\footnote{There is a classical literature on optimal patent licensing (e.g., \cite{KamienTauman1986} and \cite{KatzShapiro1986}). Those papers focus on  the effect of oligopolistic competition in the market for the patented product. Our focus is on the adverse selection problem for the patent holder.}  Earlier work assumes that cash flows are public.\footnote{ \citet[p.~666]{skrzypacz2013auctions} comments on the potential cost of verification: ``In other situations while values/costs are objective, they cannot be easily verified. For
example, in auctions selecting contractors to repair a highway, the costs of completing the project are hard to verify. On the other hand, in many commercial settings, the value of an asset/contract is at least partially observed.''}  \cite{hansen1985auctions} and \cite{riley1988ex} first observe that contingent payments can increase seller revenue relative to cash (non-contingent) payments. \cite{demarzo2005bidding} present a general auction model in which the designer specifies an ordered set of securities and agents submit security bids to the auctioneer.\footnote{They also consider ``informal'' auctions in which the auctioneer cannot commit to a rule for selecting the winning bid.} Crucially, the asset requires an upfront investment by the winning bidder. Their main finding is that ``steeper'' securities generate more revenue.\footnote{\cite{sogo2016endogenous} show that with entry costs, steeper securities can further increase revenue by attracting more bidders. By contrast, \cite{CheKim2010} observe that steeper securities can reduce revenues if each bidder's opportunity cost from acquiring the asset is increasing in the cash flow that he can generate from the asset.} In a symmetric setting, they show that a first-price auction with call options is the revenue-maximizing auction within a class of symmetric, efficient security-bid auctions. 

In our mechanism design approach, we do not impose symmetry or efficiency constraints.  \cite{inostroza2022screening} also adopt a mechanism-design approach with contingent payments, but in a setting with a single agent and public cash flows. In their optimal mechanism, as in \cite{demarzo2005bidding}, the payment received by the principal takes the form of a call option, which is flat and then linearly increasing in the income generated by the asset.\footnote{ \cite{inostroza2022screening} analyze the security received by the agent. This security is a debt contract. The payment received by the principal is therefore a call option.} By contrast, in our optimal mechanism with costly auditing, the payment to the principal is linearly increasing up to the royalty cap and then flat. \cref{sec:comparison} provides a more detailed comparison of our results with those in \cite{demarzo2005bidding} and \cite{inostroza2022screening}.

Several papers analyze optimal contingent-payment auctions under parametric restrictions on the form of the payments. In \cite{bernhardt2020costly}, payments are required to be in the form of cash plus a linear royalty at a fixed (exogenous) rate. In the setting of a takeover auction, \cite{liu2016optimal} considers equity auctions, in which each bidder's payment is an (endogenous) equity share. \cite{liu2021rent} allow for cash plus equity, and show that full extraction is possible. In all of these papers on security-bid auctions, the uninformed party designs the securities. By contrast, the finance literature on security design, beginning with \citet{nachman1994optimal}, typically assumes that the issuer has private information.\footnote{A recent exception is \citet{gershkov2023optimal}, who study optimal security design with informed, risk-averse investors.}



We model auditing as costly state verification, following
\cite{townsend1979optimal}. In models of lending to a single agent, \cite{townsend1979optimal}, \cite{diamond1984financial}, and \cite{gale1985incentive} show that it is optimal for the principal to offer a debt contract in which verification is performed only in the case of default. In those models, the agent has no private information prior to contracting. Our paper shows that the optimality of debt contracts extends to a costly verification setting with sequential screening. Moreover, we establish this optimality using the Myersonian approach. 

There are also models of costly state verification with static screening, but they focus on different issues. \cite{border1987samurai} characterize optimal tax enforcement with costly auditing. In their optimal mechanism, taxes are monotonically increasing and auditing is monotonically decreasing as a function of reported wealth. After auditing, the agent is offered a rebate if his report is revealed to be truthful. \cite{ben2014optimal}, \cite{mylovanov2017optimal}, and \cite{li2020mechanism} study optimal mechanisms for allocating a good among agents, using costly verification but no transfers. After verification, the principal can impose a limited punishment. This punishment, unlike a transfer, does not benefit the principal. 

In our model, private information arrives sequentially. At the time of contracting, each agent has an imperfect signal of his valuation, as in the single-agent models of \citet{baron1984regulation}  and \citet{courty2000sequential}, and the multi-agent model of \cite{esHo2007optimal}.\footnote{These settings are all nested by the general multi-period setting of \cite{pavan2014dynamic}. That paper provides conditions under which an envelope formula holds.} In those models, each agent learns his true valuation for an object (like a plane ticket) after contracting but \emph{before} allocation. Thus, the principal can use the allocation together with transfers to elicit the agent's new information. In our model, the new information (income) is received by the winner of the asset \emph{after} the asset is allocated. Eliciting this information requires
contingent payments.\footnote{In \cite{mezzetti2004mechanism,mezzetti2007mechanism} \emph{every} agent receives private information (his realized interdependent payoff) after the allocation. The principal can elicit this information, without verification, provided that each agent's report affects the other agents' payments, not his own.}


\section{Model} \label{sec:model}

\paragraph{Setting}
There is a principal and there are $N$ agents, labeled $i =1, \ldots, N$. The principal has an asset to allocate.   Each agent observes a private signal about his future income from acquiring the asset. Agent $i$'s signal realization, denoted $\th_i$, is called his type. Each agent $i$'s type $\th_i$ is independently drawn from a distribution $F_i$ with a continuous, strictly positive density $f_i$ on its support $\Th_i = [\ul{\th}_i, \ol{\th}_i]$. The space of type profiles is $\Th = \prod_{i=1}^{N} \Th_i$. 

If agent $i$, with type $\th_i$, receives the asset, then his income $\pi_i$ is drawn from a distribution $G_i ( \cdot |  \th_i)$ with a continuous, strictly positive density $g_i ( \cdot |  \th_i)$ over the interval $\Pi_i ( \th_i) \coloneqq [\ubar{\pi}_i (\th_i), \bar{\pi}_i ( \th_i)]$, where $0 \leq \ubar{\pi}_i (\th_i) < \bar{\pi}_i (\th_i) < \infty$.  In particular, the support of the conditional income distribution can shift with the agent's type.\footnote{With shifting supports, our regularity assumptions (imposed below) become less restrictive. Shifting supports present a few technical issues, but these can be resolved, as first observed in \cite{LiuEtal2020}.}  We assume that the functions $\ubar{\pi}_i$ and $\bar{\pi}_i$ are continuous and weakly increasing.  As in the main setting of \cite{demarzo2005bidding}, agent $i$'s income distribution depends only on his own type, not on the types of other agents. This is the analog, in our setting, of the private values assumption.\footnote{Only the winner's income is realized, so there is no need to specify the \emph{joint} distribution of $\pi_1, \ldots, \pi_N$. Thus, our private-values model implicitly allows for some dependence between different agents' realized incomes. For example, incomes could be given by $\pi_i = \th_i + \e$, for some aggregate shock $\e$ that is independent of $(\th_1, \ldots, \th_N)$.} 

Let $\Pi_i = \cup_{ \th_i \in \Th_i} \Pi_i (\th_i)$. For each $\pi_i \in \Pi_i$, let 
\begin{equation*} 
    \Th_i(\pi_i) = \Set{ \th_i \in \Th_i: \ubar{\pi}_i (\th_i) < \pi_i < \bar{\pi}_i (\th_i) }.
\end{equation*}
For each agent $i$ and each $\pi_i \in \Pi_i$, we assume that $G_i ( \pi_i | \cdot)$ is continuously differentiable over $\Th_i (\pi_i)$ and that 
\begin{equation*} 
    G_{i,2} ( \pi_i |  \th_i) < 0,
\end{equation*}
for each $\th_i$ in $\Th_i (\pi_i)$. Thus, the conditional income distribution $G_i ( \cdot | \th_i)$ is strictly increasing (in the sense of first-order stochastic dominance) in the type $\th_i$. We normalize types so that for each agent $i$, 
\begin{equation}
\label{eq:normalized}
    \E [ \pi_i |  \th_i ] = \th_i. 
\end{equation}
Thus, each agent's type equals his expected income from acquiring the asset. Finally, we assume that there exists a constant $L > 0$ such that for each agent $i$, we have $|G_{i}(\pi_i | \th_i) - G_i (\pi_i | \th_i') | \leq L |\th_i - \th_i'|$ for all $\th_i, \th_i' \in \Th_i$ and all $\pi_i \in \Pi_i$. This Lipschitz condition, together with our differentiability assumptions, allows us to differentiate under integrals.

The principal can audit the winner of the asset. This audit perfectly reveals the winner's income.\footnote{In \cref{remark:noisy}, we show that this assumption can be relaxed.} The principal's auditing cost, $c_i$, can depend on the identity $i$ of the audited agent.\footnote{We assume that the principal pays the auditing cost. In some models (e.g., \cite{diamond1984financial}), the agent pays the auditing cost. As long as $c_i$ is commonly known, our results are unaffected by who formally pays this cost since the cost can be re-allocated through transfers.}

The principal and the agents are risk-neutral. There is no discounting.\footnote{This simplifies notation. It is straightforward to incorporate a common discount rate for the principal and the agents.} The principal maximizes the expectation of payments minus auditing costs. Each agent maximizes the expectation of income minus payments.

\paragraph{Discussion} We now discuss the independent private values assumption. In design problems, it is standard to assume that types are independently distributed; otherwise, under general conditions, there exist (unrealistic) mechanisms that fully extract the surplus \citep{cremer1988full}. Our independence assumption should be interpreted as
\emph{conditional independence}, given any information that is public (or that is held by multiple agents and hence can be elicited freely). For example, in \citeapos{liu2016optimal} model of a takeover auction, the standalone value of the target firm is public, but each bidding firm privately knows its additional synergy value from acquiring the target. These synergy values are independent across bidders. Our setting admits a similar interpretation: the baseline value of the intellectual property is public, but each bidder has a private signal of his additional synergy value above this baseline.  Of course, with a single potential buyer, the independent private values assumption holds vacuously.

\paragraph{Protocol} 
    The principal commits to her strategy in the following multi-stage protocol.
\begin{enumerate}
    \item \label{it:nature} Each agent's type is realized.
    \item \label{it:bids} Each agent submits a message to the principal.
    \item \label{it:qt} The principal allocates the asset to at most one agent (called the winner) and charges each agent a \emph{transfer}.
    \item[] [If the asset is not allocated, there is no winner and the protocol ends.]
    \item \label{it:income_realized} The winner's income is realized. 
    \item \label{it:income_report} The winner submits a message to the principal. 
    \item \label{it:royalty} The principal charges the winner a \emph{royalty}. 
    \item  \label{it:a} With some probability, the principal audits the winner. 
    \item \label{it:punishments} If the winner is audited, the principal observes the winner's realized income and then charges the winner a \emph{penalty}. 
\end{enumerate}

Stages \ref{it:nature}--\ref{it:qt} are standard. If the asset is allocated, then the procedure continues. The royalty charged to the winner in stage \ref{it:royalty} can depend on the winner's message in stage \ref{it:income_report} (as well as on the winner's identity and the messages in stage \ref{it:bids}). The penalty charged to the winner in stage \ref{it:punishments}, after auditing, can be contingent on the winner's realized income (as well as on the winner's identity and the messages in stages \ref{it:bids} and \ref{it:income_report}). 

The terms \emph{royalty} and \emph{penalty} are suggestive of our motivating applications. Formally, the three kinds of payments---transfers, royalties, and penalties---are distinguished only by the information that they can depend on. This protocol is motivated by the timing used in many applications, but all payments could equivalently be delayed until the end of the game.

\paragraph{Mechanisms} By the revelation principle, there is no loss in restricting attention to direct mechanisms in which the principal elicits a type report from each agent in stage~\ref{it:bids} and an income report from the winner in stage~\ref{it:income_report}. To represent the principal's stochastic allocation decision, let \[
\mathcal{Q} = \{ (q_1, \ldots, q_N) \in \R_+^N: q_1 + \cdots + q_N \leq 1\}.
\]
In particular, the principal does not have to allocate the asset to any of the agents. If the profile $\th' = (\th_1', \ldots, \th_N')$ is reported and the asset is allocated to agent $i$, then it is sufficient for the principal to solicit from agent $i$ an income report in the set $\Pi_i ( \th_i')$. Denote the set of report histories (when agent $i$ is the winner) by
\[
    \HH_i = \big\{ (\th_1', \ldots, \th_N', \pi_i') \in \Th \times \Pi_i : \pi_i' \in \Pi_i (\th_i') \big\}.
\]

A direct mechanism for the principal specifies the following:
\begin{itemize}
	\item allocation rule $q = (q_1, \ldots, q_N) \colon \Th  \to \mathcal{Q}$;
	\item transfer rule $t = (t_1, \ldots, t_N) \colon \Th \to \R^N$;
        \item royalty rule  $r_i \colon \HH_i \to \R$, for each agent $i$;
        \item auditing rule $a_i: \HH_i \to [0,1]$, for each agent $i$;
        \item penalty rule $p_i \colon \HH_i \times \Pi_i \to \R$, for each agent $i$.
\end{itemize}

A mechanism is denoted by $(q,t,r,a,p)$, where $r = ( r_1, \ldots, r_N)$, $a = (a_1, \ldots, a_N)$, and $p = (p_1, \ldots, p_N)$. The allocation and transfer rules are standard. We describe the royalty, auditing, and penalty rules. Suppose that the type profile $\th' = (\th_1', \ldots, \th_N')$ is reported and the principal allocates the asset to agent $i$.\footnote{If $q_i ( \th') = 0$, then this history $(\th', \pi_i')$ cannot be reached. Under our definition, a mechanism specifies decisions at more histories than is necessary. This convention simplifies notation.} Since agent $i$ is the winner, the mechanism applies the rules $r_i$, $a_i$, and $p_i$. If agent $i$ reports income $\pi_i' \in \Pi_i (\th_i')$, then the principal charges agent $i$ the royalty payment $r_i ( \th', \pi_i')$. Then the principal audits agent $i$ with probability $a_i ( \th', \pi_i')$. If the audit is conducted, then agent $i$'s realized income $\pi_i \in \Pi_i$ is revealed to the principal, and the principal charges agent $i$ the penalty $p_i (\th', \pi_i', \pi_i)$.\footnote{\label{ft:message_space}If agent $i$ wins the asset after reporting $\th_i'$, then the principal accepts income \emph{reports} only from $\Pi_i (\th_i')$. But the mechanism must still specify the outcome if the agent's true income is subsequently revealed to be outside $\Pi_i (\th_i')$; such an income realization proves that the agent misreported his type (and his income).} 

Crucially, we impose a constraint on how the winner's payments depend on his realized income from the asset. Fix parameters $\phi_1, \ldots, \phi_N \in [0,1]$. 

\begin{cond}[Generalized double monotonicity] \label{L:punishment_constraint}  For each agent $i$ and each report history $(\th', \pi_i')$ in $\HH_i$, the function $p_i ( \th', \pi_i', \cdot)$ satisfies, for all $\pi_i, \hat{\pi}_i \in \Pi_i$,
\begin{equation*}
  \hat{\pi}_i > \pi_i \implies 0 \leq p_i ( \th', \pi_i', \hat{\pi}_i)  -  p_i ( \th', \pi_i', \pi_i) \leq  (\hat{\pi}_i - \pi_i) \phi_i. 
\end{equation*}
\end{cond}

Condition~\ref{L:punishment_constraint} requires that for each report history, agent $i$'s post-audit penalty is weakly increasing in his realized income, and the rate of increase is at most $\phi_i$.\footnote{The upper bound on the slope is critical; the lower bound is not. If we reduced the lower bound to $- (\hat{\pi}_i - \pi_i) C$ for any fixed constant $C > 0$, the mechanism in \cref{MAINTHEOREM} would remain optimal.  \cite{LuoYang2023} consider securities that are monotone and Lipschitz, but their focus is on coordination frictions.} Since the penalty is the only payment that depends directly on realized income, Condition~\ref{L:punishment_constraint} equivalently restricts the sensitivity of agent $i$'s total payment (summing transfers, royalties, and penalties) to his realized income. In words, if agent $i$ wins the asset, then for each additional dollar generated by the asset, agent $i$ must retain at least a fraction $1 - \phi_i$ of the dollar (and at most the full dollar).\footnote{Condition~\ref{L:punishment_constraint}  has antecedents in the literature on tax evasion. In their classic work, \cite{allingham1972income} assume an exogenous linear penalty of $ ( \pi_i - \pi_i') \phi_i$; see also \cite{kleven2011unwilling} and \cite{palonen2026avoidance}. We restrict the maximum sensitivity of the penalty, but we allow nonlinear penalties.}

If all the parameters $\phi_i$ equal $1$, then Condition~\ref{L:punishment_constraint} reduces to \emph{double monotonicity}: both maps $\pi_i \mapsto p_i (\th', \pi_i', \pi_i)$ and $\pi_i \mapsto \pi_i -  p_i (\th', \pi_i', \pi_i)$ are weakly increasing. 
Double monotonicity is a standard assumption in the security design literature; see, e.g., \cite{demarzo2005bidding}, 
\cite{inostroza2022screening}, and \cite{nachman1994optimal}.
The double-monotonicity assumption is motivated by an informal moral hazard argument. If the payment by the asset-holder ever decreases in the asset's income, then the asset-holder could generate a risk-free return by injecting cash into the project to raise measured income, thus decreasing the payment that he must make to the principal. Conversely, if the asset-holder's net profit ever decreases in the asset's income, then he could strictly benefit by burning the asset's cash flow. In many applications, the asset-holder can do better than burning money---he can divert cash for his own benefit. If the asset-holder can use fraction $1 - \phi_i$ of each dollar that he diverts, then our generalization of double monotonicity is the appropriate condition.\footnote{Condition~\ref{L:punishment_constraint} can alternatively be microfounded as follows. Suppose that an audit reveals agent $i$'s income with probability $\phi_i$ and reveals nothing otherwise. In this case, requiring the penalty to be double monotone in the winner's income (when it is revealed) is equivalent to requiring the \emph{expected} penalty to satisfy Condition~\ref{L:punishment_constraint}.} More broadly, our condition captures the moral hazard concern, noted by \cite[p.~950]{demarzo2005bidding}, that the asset-holder's incentives are dampened if he retains too small a share of the asset's income. 

In summary, the model primitives specify, for each agent $i$, the type distribution $F_i$, conditional income distribution $G_i$, auditing cost $c_i$, and maximum penalty sensitivity $\phi_i$. 


\section{Principal's program} \label{sec:program}

We now formulate the principal's optimization problem. To state the incentive constraints, we introduce additional notation. For each allocation rule $q_i$ and transfer rule $t_i$, define the associated interim rules by 
\begin{equation*}
    Q_i ( \th_i') = \E_{\th_{-i}} \Brac{ q_i (\th_i', \th_{-i})},
    \qquad
    T_i(\th_i') = \E_{\th_{-i}} \Brac{t_i(\th_i', \th_{-i})}.
\end{equation*}
Consider a mechanism $(q,t,r,a,p)$. Fix an agent $i$. For any type report $\th_i' \in \Th_i$, income report $\pi_i' \in \Pi_i (\th_i')$, and true income $\pi_i \in \Pi_i$, define the utility
\begin{multline} \label{eq:def_ui}
    u_i (\th_i', \pi_i'| \pi_i) \\ = \E_{\th_{-i}} \Big[q_i (\th_i', \th_{-i}) \big( \pi_i - r_i(\th_i', \th_{-i},\pi_i')    - a_i(\th_i', \th_{-i}, \pi_i') p_i (\th_i', \th_{-i}, \pi_i', \pi_i)\big)\Big].
\end{multline}
If $Q_i ( \th_i') = 0$, then $u_i (\th_i', \pi_i' | \pi_i) = 0$. If $Q_i (\th_i') > 0$, then $u_i (\th_i', \pi_i' | \pi_i)/Q_i (\th_i')$ is agent $i$'s  expected gross utility (excluding upfront transfers) conditional upon reporting $\th_i'$, winning the asset, privately observing income $\pi_i$, and then reporting income $\pi_i'$. This utility term $u_i (\th_i', \pi_i'| \pi_i)$ allows us to conveniently express agent $i$'s income-reporting incentive constraint without conditioning on zero-probability events. 

There is one subtlety introduced by shifting supports. If type $\th_i$ reports $\th_i'$ and $\Pi_i (\th_i') \not\supseteq \Pi_i ( \th_i)$, then the realized income $\pi_i$ may be outside $\Pi_i(\th_i')$. In this case, it is not feasible for agent $i$ to report his income truthfully.  We introduce notation for reporting ``as truthfully as possible.'' For any type report $\th_i' \in \Th_i$ and any income $\pi_i \in \Pi_i$, let $\proj_{\Pi_i (\th_i')} \pi_i$ denote the element in $\Pi_i (\th_i')$ that is closest to $\pi_i$. In particular, $\proj_{\Pi_i(\th_i')} \pi_i = \pi_i$ if $\pi_i$ is in $\Pi_i (
\th_i')$. For any types $\th_i, \th_i' \in \Th_i$, let
\begin{equation} \label{eq:Ui}
    U_i(\th_i' | \th_i) =  \E_{\pi_i|\th_i}  \Brac{ u_i(\th_i', \proj_{\Pi_i(\th_i')} \pi_i | \pi_i) }.
\end{equation}
This expectation is taken with respect to the true income distribution $G_i ( \cdot | \th_i)$. Thus, $U_i(\th_i' | \th_i)$ is the expected gross utility (excluding upfront transfers) for type $\th_i$ if he reports type $\th_i'$ and then reports his income as truthfully as possible whenever he wins the asset.

We consider the principal's problem of choosing a direct mechanism $(q,t,r,a,p)$ satisfying Condition~\ref{L:punishment_constraint} to solve:
    \begin{align*}
   \max \, \sum_{i=1}^{N}  \E_{\th} \left[t_i(\th) + q_i(\th) \E_{\pi_i | \th_i} \Brac{ r_i(\th,\pi_i) + a_i(\th,\pi_i) (p_i ( \th, \pi_i,  \pi_i) - c_i)}  \right]
    \end{align*}
    subject to the following constraints, for each agent $i$:
    \begin{align}
        \tag{$\mathrm{IC}_\pi$} \label{IC2}
    &u_i (\th_i, \pi_i | \pi_i)  \geq u_i (\th_i, \pi_i' | \pi_i) , \quad \th_i \in \Th_i, \quad \pi_i, \pi_i' \in \Pi_i (\th_i) \\
     \tag{$\mathrm{IC}_\th$} \label{IC1}
       &U_i(\th_i |\th_i)   - T_i(\th_i) \geq U_i(\th_i'| \th_i)   - T_i(\th_i'), \quad \th_i, \th_i' \in \Th_i \\
    \tag{IR} \label{IR}
    &U_i(\th_i | \th_i)   - T_i(\th_i) \geq 0, \quad \th_i \in \Th_i.
    \end{align}

In the principal's objective, the $i$-th term in the summation is the expected payment made by agent $i$ (including transfers, royalties, and penalties) net of the principal's expected cost of auditing agent $i$. Technically, some of the constraints are relaxations of the full incentive constraints, as we describe below. We will solve this relaxed problem and then confirm that our solution satisfies the full incentive constraints. 
    
The constraint \eqref{IC2} captures incentive compatibility for the winner of the asset at the income-reporting stage.\footnote{In particular, \eqref{IC2} takes into account agent $i$'s inference (about the other agents' type reports) from winning the asset. The constraint \eqref{IC2} does not take into account agent $i$'s further inference from the upfront transfer that he is charged. Thus, \eqref{IC2} is technically a relaxation of the income-reporting incentive constraint. In our solution of the relaxed problem, the reporting, auditing, and penalty rules for the winner are all independent of the other agents' type reports. Thus, the winner does not benefit from learning the others' type reports.} Suppose that type $\th_i$ reports truthfully. If $Q_i ( \th_i) > 0$, then there is a positive probability that agent $i$ wins the asset. In this case, the constraint  \eqref{IC2} ensures that if agent $i$ wins the asset, it is optimal for him to truthfully report his income realization, whatever its value. If $Q_i(\th_i) = 0$, then \eqref{IC2} automatically holds---both sides equal zero by the definition of $u_i$.\footnote{A similar approach is followed in the definition of Bayes correlated equilibrium \citep[Definition 1, p.~493]{BergemannMorris2016}.} 

The constraint \eqref{IC1} says that every type $\th_i$ weakly prefers (a) truthfully reporting his type and then truthfully reporting his subsequent income if he wins the asset, to (b) misreporting his type and then reporting his income ``as truthfully as possible'' if he wins the asset. This is a relaxation of the full incentive-compatibility constraint, which requires that all double deviations are unprofitable.\footnote{This relaxation can be strict only with shifting supports. With common support, \eqref{IC2} implies that it is optimal for the winner to report his income truthfully, even after misreporting his type.}
        
Finally, \eqref{IR} ensures that every type weakly prefers participating in the mechanism to his outside option of zero utility. 

\section{Benchmarks} \label{sec:benchmarks}

In this section, we describe the optimal mechanism in two extreme cases: non-contingent payments and fully contingent payments.

\paragraph{Non-contingent payments} Suppose that the principal cannot charge payments that are contingent on the winner's realized income, i.e., $\phi_i = 0$ for all $i$.\footnote{This benchmark also corresponds to the limit as auditing becomes prohibitively costly.} After the asset is allocated, the principal can still ask the winner to report his income. But without the threat of contingent payments, the winner will make whichever income report minimizes his expected payment. Thus, incentive compatibility implies that the winner's expected post-allocation payment cannot depend on his reported income. Therefore, expected payments can depend only on the agents' initial type reports, and the model reduces to the classical cash auction setting of \cite{myerson1981optimal}. Agent $i$'s valuation for the asset is $\E[\pi_i | \th_i]$, which equals $\th_i$ by our normalization. The optimal cash auction allocates the asset according to each agent $i$'s Myerson virtual value
\[
    \psi_i^{M} ( \th_i) = \th_i - \frac{1 - F_i (\th_i)}{f_i(\th_i)}.
\]
Transfers are pinned down by the envelope theorem. In the main model with verification and penalties, this cash auction is always feasible, but it is generally suboptimal.

\paragraph{Fully contingent payments} Suppose that the winner's income from the asset is publicly revealed and that the principal can charge the winner a payment that depends arbitrarily on the realized income. In this case, the principal can fully extract the surplus using the following modified first-price auction. The allocation and upfront transfers are the same as in a standard first-price auction. After allocating the asset, the principal charges the winner a penalty equal to the difference between his realized income and his earlier bid. 

In this auction, no matter how the agents bid, the winner's total payment equals his realized income from the asset, and the losers pay nothing. Each agent's expected utility is zero, and hence every strategy profile is a dominant-strategy equilibrium. In the truthful equilibrium, the asset is allocated to the agent with the highest type, so the principal extracts the full surplus $\E \Brac{ \max \{ 0, \th_1, \ldots, \th_N\}}$. This full-extraction mechanism is feasible in the main model if auditing is costless ($c_i = 0$ for all $i$) and  $\phi_i = 1$ for each $i$.\footnote{Similarly, \cite{cremer1987} observes that in the setting of \cite{hansen1985auctions}, full extraction is possible if the limited liability constraint is dropped.} 

If penalties can depend arbitrarily on realized income, then even if auditing is costly, the principal can approximate full extraction by adjusting the modified first-price auction as follows. For any $\e > 0$, the principal can audit the winner with probability $\e$ and then scale the penalty rule by $1/\e$. Expected payments are unchanged, and the auditing cost vanishes in the limit as $\e$ tends to $0$.  In our model, \cref{L:punishment_constraint} rules out such mechanisms. 


\section{Solving for an optimal mechanism} \label{sec:main}

In this section, we use the Myersonian approach to solve for an optimal mechanism. Then we analyze comparative statics.

\subsection{Myersonian approach}

We use the envelope theorem to obtain an upper bound on the principal's objective. This bound depends only on the allocation and auditing rules, not on any of the payment rules. First, we introduce additional notation. For each $\th_i \in \Th_i$ and each $\pi_i \in (\ubar{\pi}_i (\th_i), \bar{\pi}_i (\th_i))$, let
\[
    \mu_i (\th_i, \pi_i) = - \frac{G_{i,2} ( \pi_i | \th_i)}{g_i(\pi_i | \th_i)} \cdot \frac{1 - F_i (\th_i)}{f_i(\th_i)}.
\]
Since the partial derivative $G_{i,2}$ is strictly negative, we have $\mu_i(\th_i, \pi_i) > 0$ if $\th_i < \bar{\th}_i$, and we have $\mu_i(\bar{\th}_i, \pi_i) = 0$. 


\begin{lem}[Payoff bound] \label{OBJECTIVELEMMA}
    Let $(q,t,r,a,p)$ be a mechanism that satisfies Condition~\ref{L:punishment_constraint} and the constraints \eqref{IC2}, \eqref{IC1}, and \eqref{IR} for each agent $i$. The principal's expected payoff from $(q,t,r,a,p)$ is at most
    \begin{equation*} 
    \E_\th \Brac{ \sum_{i=1}^{N} q_i (\th) \Psi_i (\th)},
    \end{equation*}
    where for each agent $i$, the function $\Psi_i \colon \Th \to \R$ is given by 
    \[
        \Psi_i (\th) = \th_i - \frac{1  -F_i(\th_i)}{f_i(\th_i)} + \E_{\pi_i | \th_i} \Brac{ a_i (\th, \pi_i) ( \mu_i (\th_i, \pi_i) \phi_i - c_i )}.
    \]
\end{lem}


To prove
\cref{OBJECTIVELEMMA}, we apply the envelope theorem to the income-reporting and type-reporting incentive constraints. The difficulty is that the penalty can depend on the winner's \emph{realized} income, so the allocation rule does not pin down the agents' information rents. Using Condition~\ref{L:punishment_constraint}, we derive a lower bound on each agent's information rent, which yields an upper bound on the principal's payoff. Below, we will find a mechanism that achieves this upper bound and hence is optimal. 

In \cref{OBJECTIVELEMMA}, the coefficient $\Psi_i (\th)$ is an \emph{endogenous} virtual value that depends on the full type profile $\th$ and on the principal's auditing rule. Note that $\Psi_i (\th)$ is the sum of the Myerson virtual value $\psi_i^M (\th_i)$ and an additional expectation term, which we call the \emph{auditing term}. This term takes a simple form in two extreme cases. If the principal never audits agent $i$, then the auditing term vanishes and we get $\Psi_i(\th) = \psi_i^M (\th_i)$. If the principal always audits agent $i$, then it can be checked that 
\begin{equation*}
    \Psi_i ( \th) = \th_i - (1 - \phi_i)\frac{1  -F_i(\th_i)}{f_i(\th_i)}  - c_i.
\end{equation*}
The sensitivity parameter $\phi_i$ controls the share of agent $i$'s information rent that the principal can recover through penalties, upon auditing the agent. 

In the auditing term, the expression inside the expectation reflects the direct cost $c_i$ of auditing agent $i$ and the indirect benefit from reducing agent $i$'s information rent. Here is a heuristic derivation. Auditing agent $i$ at history $(\th, \pi_i)$ allows the principal to increase the slope of the royalty payment at that history by at most $\phi_i$. This raises the royalty paid by type $\th_i$ in the event 
that his realized income is at least $\pi_i$. The probability of this event is $1 - G_i (\pi_i | \th_i)$. The sensitivity of this probability to agent $i$'s true type is $-G_{i,2} (\pi_i | \th_i)$. Therefore, by the envelope theorem, auditing at history $(\th, \pi_i)$ reduces by at most $-G_{i,2} (\pi_i | \th_i)\phi_i q_i (\th)$ the information rent of every type above $\th_i$. The mass of such types is $1 - F_i (\th_i)$, so the expected information rent is reduced by at most
\[
    -G_{i,2} (\pi_i | \th_i)  (1 - F_i (\th_i) ) \phi_i q_i(\th).
\]
In the auditing term, this coefficient on $q_i(\th)$ is divided by the density $g_i (\pi_i | \th_i) f_i(\th_i)$ to offset the scaling from the expectation over $(\th_i, \pi_i)$. 


\subsection{Optimal mechanism}

To solve for an optimal mechanism, we find allocation and auditing rules that pointwise maximize the expression inside the expectation in the payoff bound (\cref{OBJECTIVELEMMA}). Under suitable regularity conditions, we show that this mechanism satisfies global incentive compatibility. 

The endogenous virtual value, $\Psi_i (\th)$, depends on the principal's auditing rule. To maximize $\Psi_i (\th)$, the principal audits agent $i$ if and only if the resulting reduction in information rent outweighs the direct cost of the audit: $\mu_i (\th_i, \pi_i) \phi_i \geq c_i$. With this optimal auditing rule, the endogenous virtual value $\Psi_i(\th)$ becomes $\psi_i (\th_i)$, where
\begin{equation*}
        \psi_i(\th_i) = \th_i - \frac{1-F_i(\th_i)}{f_i(\th_i)} + \E_{\pi_i|\th_i} \left[ \max \{\mu_i(\th_i,\pi_i) \phi_i -c_i, 0 \} \right].
\end{equation*}
Hereafter, we call $\psi_i (\th_i)$ the \emph{virtual value} of type $\th_i$.  Unlike $\Psi_i (\th)$, the virtual value $\psi_i(\th_i)$ depends only on agent $i$'s type (not on the principal's auditing rule or on the types of the other agents).\footnote{\cite{liu2016optimal} derives an expression for the virtual value in equity auctions. 
The exact formula is different from ours, but both expressions reflect the fact that contingent payments reduce information rents. \cite{ball2019probabilistic} also derive an expression for the virtual value in their model of probabilistic verification. Before the allocation, the principal can generate a binary, noisy signal of whether the agent is misreporting, and the principal can exclude the agent if a misreport is detected.}

In order to solve for an optimal mechanism, we introduce suitable regularity conditions. Consider a real-valued function $h$ whose domain, $\dom h$, is a convex subset of $\R$. We say that $h$ is \emph{single-crossing from above} if for all $x,x' \in \dom h$ with $x < x'$, we have: $h(x) \mathrel{\leq(<)} 0 \implies h(x') \mathrel{\leq(<)} 0$. 


\begin{as}[Regularity] \label{AA} For each agent $i$, the following hold:
\begin{enumerate}[label = (\alph*), ref = \alph*]
    \item \label{it:single-crossing} $\mu_i ( \th_i, \pi_i) \phi_i - c_i$ is single-crossing from above in $\th_i$ and in $\pi_i$;
    \item \label{it:regularity} $\psi_i (\th_i)$ is strictly increasing in $\th_i$.
\end{enumerate}
\end{as}


\cref{AA}.\ref{it:single-crossing} ensures that the pointwise optimal  rule for auditing agent $i$ is weakly decreasing in agent $i$'s type and income. \cref{AA}.\ref{it:regularity} is the analog of Myerson regularity. It ensures that the virtual surplus–maximizing allocation rule for agent $i$ is weakly increasing in agent $i$'s type, for each fixed type profile of the other agents. \cref{AA} is strictly weaker than the regularity assumptions in \cite{esHo2007optimal}, as we show in Appendix \ref{A:SHIFT}. 


To describe our optimal mechanism, we introduce additional notation. Using \cref{AA}.\ref{it:single-crossing}, we define for each agent $i$ a weakly decreasing function $\pi_i^\ast \colon \Th_i \to [0, \infty]$ with the property that for all types $\th_i \in \Th_i$ and all  $\pi_i \in (\ubar{\pi}_i (\th_i), \bar{\pi}_i (\th_i))$, we have $\pi_i \leq \pi_i^\ast (\th_i) \iff \mu_i (\th_i, \pi_i) \phi_i \geq c_i$.\footnote{For each $\th_i \in \Th_i$, let $\Pi_i^\ast(\th_i) = \{ \pi_i \in (\ubar{\pi}_i (\th_i), \bar{\pi}_i (\th_i)) : \mu_i ( \th_i, \pi_i) \phi_i \geq c_i \}$. Define $\pi_i^\ast$ by 
\[
    \pi_i^\ast (\th_i)
    = 
    \begin{cases}
        \infty &\text{if}~\Pi_i^\ast (\th_i) = (\ubar{\pi}_i (\th_i), \bar{\pi}_i (\th_i)),\\
     \sup \Pi_i^\ast( \th_i)  &\text{if}~     \varnothing \subsetneq   \Pi_i^\ast (\th_i) \subsetneq (\ubar{\pi}_i (\th_i), \bar{\pi}_i (\th_i)), \\
       0 &\text{if}~\Pi_i^\ast (\th_i) = \varnothing.
    \end{cases}
\]
} Let
\begin{equation} \label{eq:Psi}
\Phi_i(\th_i) 
=  -\phi_i \int_{0}^{\pi_i^\ast ( \th_i)} G_{i,2}(\pi_i|\th_i) \de \pi_i.
\end{equation}
The normalization \eqref{eq:normalized} implies that  $- G_{i,2} (\cdot | \th_i)$ integrates to $1$. Since $- G_{i,2} (\cdot | \th_i)$ is nonnegative, we have $0 \leq \Phi_i (\th_i) \leq \phi_i$. To describe our optimal mechanism, let $[ \cdot ]$ denote the indicator function for the predicate it encloses.

\begin{thm}[Optimal mechanism] \label{MAINTHEOREM}
       Under \cref{AA}, the following mechanism $(q^\ast, t^\ast, r^\ast, a^\ast, p^\ast)$ is optimal. For each agent $i$, the allocation and auditing rules are given by\footnote{To cover the case $N = 1$, we adopt the convention that the maximum of the empty set is $-\infty$.}
        \[
            q_i^\ast ( \th) = [ \psi_i (\th_i) \geq 0~\text{and}~ \psi_i (\th_i) > \max_{j \neq i} \psi_j (\th_j)],  \qquad 
            a_i^\ast ( \th, \pi_i') = [ \pi_i' < \pi_i^\ast ( \th_i)].
        \]
        The royalties and penalties are given by
        \[
            r_i^\ast ( \th, \pi_i') =  \min \{  \pi_i', \pi_i^\ast ( \th_i) \} \phi_i,
            \qquad
            p_i^\ast ( \th, \pi_i', \pi_i) = 
            \min\{ \pi_i, \pi_i^\ast ( \th_i)\} \phi_i - \min\{ \pi_i', \pi_i^\ast ( \th_i)\} \phi_i.
        \]
        The upfront transfers are given by
        \[
        t_i^\ast(\th) = q_i^\ast(\th) \E_{\pi_i|\th_i} \Brac{ \pi_i - r_i^\ast(\th,\pi_i) } -  \int_{\ul{\th}_i}^{\th_i} q_i^\ast(z_i,\th_{-i})(1-\Phi_i(z_i))  \de z_i.
        \]
        Moreover, $(q^\ast, t^\ast, r^\ast, a^\ast, p^\ast)$ is dominant-strategy incentive compatible and dominant-strategy individually rational.
\end{thm}


We discuss the components of the mechanism $(q^\ast, t^\ast, r^\ast, a^\ast, p^\ast)$ in turn. The asset is allocated to the agent with the highest virtual value, provided that this virtual value is nonnegative.\footnote{Since $\psi_i$ is strictly increasing, ties between agents occur with probability zero. To keep notation simple, we assume that the asset is not allocated in the case of a tie.} Otherwise, the asset is not allocated. The allocation probability $q_i^\ast$ is weakly increasing in agent $i$'s type $\th_i$ because $\psi_i$ is strictly increasing (by \cref{AA}.\ref{it:regularity}). The virtual value $\psi_i (\th_i)$ is weakly greater than the Myerson virtual value $\psi_i^M ( \th_i)$. Relative to Myerson's optimal cash auction,  the asset is allocated at more type profiles (in the sense of weak set inclusion). 

Next, we describe the auditing, royalty, and penalty rules. Agent $i$'s type report $\th_i$ determines his \emph{royalty cap} $\pi_i^\ast (\th_i) \phi_i$. Suppose that agent $i$ wins the asset after reporting type $\th_i$. If agent $i$ subsequently reports income of at least $\pi_i^\ast (\th_i)$, then he pays a royalty equal to the royalty cap $\pi_i^\ast (\th_i) \phi_i$ and he is not audited. If he instead reports income below $\pi_i^\ast ( \th_i)$, then he pays a royalty equal to the fraction $\phi_i$ of his reported income, and he is subsequently audited. If the audit reveals that his true income is different from his reported income, then he is charged a penalty equal to the underpaid royalties (or he is given a refund for the overpaid royalties).\footnote{\label{ft:refund}Refunding the agent for overpaid royalties is not necessary. It would also be optimal to use the penalty function $\bar{p}_i(\th,\pi_i',\pi_i)= \max\{ p_i^\ast ( \th,\pi_i',\pi_i),  0\}$. More generally, $p_i^\ast$ can be replaced with any feasible penalty function that agrees with $p_i^\ast$ on path and is weakly larger than $p_i^\ast$ off path.} This penalty format is consistent with the medical device royalty agreement described in the introduction.

The royalty cap $\pi_i^\ast (\th_i) \phi_i$ is weakly decreasing in agent $i$'s type report $\th_i$; we show this in the proof of \cref{MAINTHEOREM}. If agent $i$ reports a higher type, then upon winning the asset, he faces a lower royalty cap and hence is less likely to be audited. From a statistical standpoint, this auditing rule may be counterintuitive. When an agent's true type is higher, a low income report is \emph{more suspicious}. But in equilibrium, all reports are truthful. The optimal auditing rule prescribes the minimal level of auditing that preserves the incentive compatibility of the associated royalty rule. In turn, the royalty rule optimally screens the agents at the type-reporting stage according to their beliefs about future income. Higher type reports result in lower royalty caps, which reward high income realizations. These high realizations are more likely for higher types. In particular, the highest type, $\bar{\th}_i$, is never audited (provided that $c_i > 0$), and the virtual value $\psi_i (\bar{\th}_i)$ equals the true value $\bar{\th}_i$. Allocating to type $\bar{\th}_i$ does not increase information rent for higher types, so there is no motive to downward distort or audit. 

Finally, the interim transfers are pinned down by incentive compatibility. The information rent for type $\th_i$ is given by
\[
       \int_{\ul{\th}_i}^{\th_i} Q_i^\ast (z_i) \Paren{ 1 - \Phi_i(z_i)} \de z_i.
\]
This expression is similar to that in a cash auction, but here the allocation probability in the integral is scaled by the factor $1 - \Phi_i$. From \eqref{eq:Psi}, we see that for each report $\th_i$, the term $\Phi_i(\th_i)$ is between $0$ (when no income is subject to royalties) and $\phi_i$ (when all income is subject to royalties at rate $\phi_i$). 

Bayesian incentive compatibility depends only on interim payments. The particular payment rules in \cref{MAINTHEOREM}  are chosen for their special properties. Transfers and royalty payments are always nonnegative. Equilibrium penalties are zero. Agent $i$ pays an upfront transfer only if he wins the asset. Moreover, the mechanism $(q^\ast, t^\ast, r^\ast, a^\ast, p^\ast)$ is dominant-strategy incentive compatible and dominant-strategy individually rational. Whatever agent $i$ believes about his opponents' reports, it is optimal for him to participate in the mechanism and then report truthfully at every stage. Since $q_i^\ast (\th_i, \th_{-i})$ is weakly increasing in $\th_i$ and $\pi_i^\ast (\th_i)$ is weakly decreasing in $\th_i$, it follows from dominant-strategy incentive compatibility that $t_i^\ast ( \th_i, \th_{-i})$ is weakly increasing in $\th_i$.

\begin{rem}[Noisy auditing] \label{remark:noisy} We have assumed that an audit perfectly reveals the winner's income. If the auditing technology instead generates a noisy, unbiased signal of the winner's income, then we show that the principal's optimal payoff is unchanged. Formally, suppose that an audit of the winning agent $i$ generates a random signal $\z_i$, where $\E [ \z_i | \pi_i] = \pi_i$. Assume that the conditional signal distribution is weakly increasing in $\pi_i$ (with respect to  first-order stochastic dominance). We impose the analog of \cref{L:punishment_constraint}, with $\z_i$ in place of $\pi_i$. We show that the principal's optimal payoff is the same as in the main model.  This payoff is achieved by using essentially the same mechanism as in \cref{MAINTHEOREM}, except that the penalty rule is $\tilde{p}_i ( \th, \pi_i', \z_i) = (\z_i - \pi_i') \phi_i$. Thus, $\E_{\z_i | \pi_i} [\tilde{p}_i ( \th, \pi_i', \z_i)] = (\pi_i - \pi_i') \phi_i$. For the proof, see \cref{sec:noisy_auditing}.
\end{rem}


We next introduce a general class of distributions, termed \emph{transformed additive noise}, that satisfy \cref{AA}.\footnote{This class generalizes many of the examples in \cite{esHo2007optimal}. Note that Example 3 in \citet[p.~712]{esHo2007optimal} contains an algebraic error: the correct mixed partial $u_{i12}$ is nonnegative. This example must be modified to satisfy their regularity conditions.} It is convenient to first introduce these distributions in terms of an unnormalized signal, $\g_i$, for agent $i$. Then we normalize agent $i$'s signal by setting $\th_i = \E[ \pi_i | \g_i]$. Suppose that for each agent $i$, income is given by
\[
    \pi_i = \a_i ( \g_i + \e_i),
\]
where
\begin{itemize}
    \item $\g_i$ is agent $i$'s (unnormalized) signal, which has a continuous, positive density and a weakly increasing hazard rate over its support $[\ubar{\g}_i, \bar{\g}_i]$; 
    \item $\e_i$ is a random variable, independent of $\g = (\g_1, \ldots, \g_N)$, with continuous, positive density over its support $[\ubar{\e}_i, \bar{\e}_i]$; 
    \item $\a_i \colon [ \ubar{\g}_i + \ubar{\e}_i, \bar{\g}_i + \bar{\e}_i] \to [0, \infty)$ is a twice continuously differentiable function satisfying $\a_i'(x) > 0$ and $\a_i''(x) \leq 0$ for all $x \in [ \ubar{\g}_i + \ubar{\e}_i, \bar{\g}_i + \bar{\e}_i]$. 
\end{itemize}
With the renormalization $\th_i = \E [ \pi_i | \g_i]$ for each agent $i$, \cref{AA} is satisfied; see \cref{sec:proof_transformed_additive} for a proof. In the special case that $\a_i$ is the identity function and $\e_i$ has mean zero, no normalization is required, and we get $\pi_i = \th_i + \e_i$. We call this special case the \emph{additive noise} specification. Under the additive noise specification, it can be verified that $\mu_i (\th_i, \pi_i)$ depends only on $\th_i$, not on $\pi_i$. 

\begin{figure}
\centering
\begin{tikzpicture}
\begin{axis}[
  width=0.78\linewidth,
  xmin=0.95, xmax=1.50,
  ymin=0.95, ymax=1.32,
  axis lines=left,
  axis line style={-Stealth},
  clip=false,
  xlabel={$\pi_i$},
  xlabel style={at={(axis description cs:1,0)}, anchor=west, font=\small},
  xtick={
    1.009950, 
    1.048809, 
    1.125000, 
    1.225000, 
    1.421267, 
    1.449138  
  },
  xticklabels={
    {\hspace{-8pt}$\underline{\pi}_i(\theta_i')$},
    {\hspace{8pt}$\underline{\pi}_i(\theta_i'')$},
    {$\pi_i^*(\theta_i'')$},
    {$\pi_i^*(\theta_i')$},
    {\hspace{-12pt}$\overline{\pi}_i(\theta_i')$},  
    {\hspace{12pt}$\overline{\pi}_i(\theta_i'')$}   
  },
  xticklabel style={
    font=\scriptsize,
    rotate=0,        
    anchor=north,    
    inner xsep=1pt
  },
  ytick={1,1.3},
  yticklabel style={font=\scriptsize},
  ticklabel style={font=\small},
]

\addplot[dotted] coordinates {(0.95, 0.95) (1.32,1.32)} node[pos =1 , right] {$\pi_i$};

\addplot[very thick, Blue] coordinates {
  (1.009950,1.046829)
  (1.225000,1.261878)
  (1.421267,1.261878)
}
node[pos=1,right,Blue]
{$t_i^\ast(\theta_{i}', \ubar{\th}_{-i}) + r_i^\ast(\theta_{i}', \ubar{\th}_{-i},\pi_i)$};

\addplot[very thick, Orange] coordinates {
  (1.048809,1.106001)
  (1.125000,1.182192)
  (1.449138,1.182192)
}
node[pos=1,right,Orange]
{$t_i^\ast(\theta_{i}'', \ubar{\th}_{-i}) + r_i^\ast(\theta_{i}'', \ubar{\th}_{-i},\pi_i)$};




\end{axis}
\end{tikzpicture}
\caption{
Winner's total equilibrium payment as a function of income for two type reports. }
       \label{fig:roy}
\end{figure}

\cref{fig:roy} illustrates the mechanism from \cref{MAINTHEOREM} in an example with transformed additive noise. Suppose that the $N$ agents are symmetric. For each agent $i$, let $\pi_i= \sqrt{\g_i+\varepsilon_i}$, where agent $i$'s unnormalized signal $\g_i$ is uniformly distributed over $[1,2]$, and $\e_i$ is independently and uniformly distributed over $[0,1]$. Agent $i$'s normalized type is given by $\th_i(\g_i) = \E[\pi_i | \g_i]$. In this case, we have $\mu_i(\th_i(\g_i),\pi_i) = (2-\g_i)/(2\pi_i)$. Let $\phi_i = 1$ and $c_i = 0.4$. Given two types  $\th_i'$ and $\th_i''$ with $\th_i' < \th_i''$, we consider the case in which agent $i$'s opponents all report the lowest possible type.\footnote{This vector of opponent reports is denoted by $\ubar{\th}_{-i}$. Here, $\Th_i \approx [1.22,1.58]$ for all $i$. The two types in the plot are $\th_i' \approx 1.23$ and $\th_i'' \approx 1.26$ (corresponding to the unnormalized signals $\g_i'=1.02$ and $\g_i''=1.10$). The associated income supports are $\Pi_i(\th_i')\approx [1.01,1.42]$ and $\Pi_i(\th_i'')\approx [1.05,1.45]$, and the auditing thresholds are $\pi_i^*(\th_i') \approx 1.23$ and $\pi_i^*(\th_i'') \approx 1.13$. In this example, types above approximately $1.28$ (equivalently, unnormalized signals above approximately $1.14$) are never audited.} The plot is similar for other opponent reports, provided that agent $i$ still wins the asset.\footnote{More precisely, with a higher vector of opponent reports, each equilibrium payment curve is translated upward.} For each of these two types, \cref{fig:roy} plots agent $i$'s total equilibrium payment as a function of his income realization. Equilibrium penalties are zero, so they are omitted from the formulas. 

In \cref{fig:roy}, the type-$\th_i''$ payment curve crosses the type-$\th_i'$ payment curve from above. That is, the higher type makes higher payments at low income realizations and lower payments at high income realizations. This crossing property is quite general; it follows from dominant-strategy incentive compatibility, as we show in the proof of \cref{MAINTHEOREM}. Agent $i$'s realized utility is the difference between $\pi_i$ (shown as the 45-degree line) and the payment curve. For both types $\th_i'$ and $\th_i''$, the realized utility is negative for some low income realizations and positive for some high income realizations. Individual rationality ensures that for each type, the expected difference between $\pi_i$ and the payment curve is nonnegative, where the expectation is with respect to the income distribution for that type.

\cref{fig:roy} highlights that each equilibrium payment resembles a debt contract. The slope before the cap is $\phi_i$ because 
generalized double monotonicity (Condition~\ref{L:punishment_constraint}) binds.  In a debt contract, the corresponding slope is $1$ because the standard double-monotonicity constraint binds. The flat section beyond the cap, like that in a debt contract, saves on auditing costs. Our model shows that the intuition from the static, single-agent model of \cite{townsend1979optimal} extends to a richer setting with sequential screening. By contrast, in the optimal security-bid auction in \cite{demarzo2005bidding}, the equilibrium payments are call options, which are flat up to a threshold, and then increasing with slope $1$; for a more detailed comparison with their results, see \cref{sec:comparison}.

\subsection{Special cases}

We consider two special cases in which the optimal mechanism in \cref{MAINTHEOREM} takes a simple form.

First, suppose that auditing is free (i.e., $c_i = 0$ for all $i$). In this case, it is optimal for the principal to always audit the winner.  The virtual value $\psi_i (\th_i)$ becomes
    \[
        \psi^0_i (\th_i) = \th_i - (1 - \phi_i) \frac{1 - F_i(\th_i)}{f_i(\th_i)}.
    \]
Compared with Myerson's virtual value, the information rent term is scaled by $1- \phi_i$, because the fraction $\phi_i$ of the information rent is extracted through royalty payments. In particular, if $\phi_i =1$ for all $i$, then the virtual value coincides with the true value, and the principal achieves full extraction, consistent with the benchmark in \cref{sec:benchmarks}.

With free auditing, \cref{AA} is satisfied as long as $\psi_i^0$ is strictly increasing. This holds, in particular, if the type distribution $F_i$ has a weakly increasing hazard rate. Recall that $[ \cdot ]$ denotes the indicator function for the predicate it encloses. The following corollary is immediate as a special case of
\cref{MAINTHEOREM}. 

    \begin{cor}[Free auditing] \label{CORFREEA} Suppose that $c_i = 0$ for all $i$. If $\psi^0_i$ is strictly increasing for each agent $i$, then the mechanism $(q^\ast, t^\ast, r^\ast, a^\ast, p^\ast)$ from \cref{MAINTHEOREM} is optimal, and takes the following form: 
    \begin{equation*}
    \begin{aligned}
       q_i^\ast (\th) &=  [ \psi_i^0 (\th_i) \geq 0~\text{and}~ \psi_i^0 (\th_i) > \max_{j \neq i} \psi_j^0 (\th_j)], \\
       t_i^\ast ( \th) &= (1 - \phi_i) \Paren{ q_i^\ast (\th) \th_i - \int_{\ul{\th}_i}^{\th_i} q_i^\ast (z_i, \th_{-i}) \de z_i},
    \end{aligned}
    \end{equation*}
    and
    \[
   a_i^\ast (\th, \pi_i') = 1, \qquad r_i^\ast (\th, \pi_i') = \pi_i' \phi_i , \qquad p_i^\ast (\th, \pi_i', \pi_i) =  (\pi_i - \pi_i') \phi_i.
    \]
    \end{cor} 

Here, the winner pays a linear royalty at rate $\phi_i$. \cite{bernhardt2020costly} derive the same virtual value $\psi_i^0(\th_i)$ in their analysis of optimal auctions in which payments must be in the form of cash plus a linear royalty at a fixed (exogenous) rate.\footnote{Also, in \citeapos{liu2016optimal} optimal equity-auction model, in the limit that synergies are much smaller than standalone values, the virtual value converges to $\psi_i^0(\th_i)$, with $\phi_i = V_i / (V_i + V_T)$, where $V_i$ is firm $i$'s standalone value and $V_T$ is the standalone value of the target \citep[p.~110, footnote 15]{liu2016optimal}.} Their model features entry costs, so they must take into account which types of each bidder will participate. 
    
For the second special case, suppose that there is only one potential buyer. For this case, we drop agent subscripts.  With one buyer, the solution from \cref{MAINTHEOREM} can be implemented by offering a menu of contracts. In this optimal menu, each contract specifies linear royalties at rate $\phi$ up to a royalty cap, with auditing if and only if the agent claims to owe less than the royalty cap. Each contract in the menu is distinguished by its upfront price $t^\ast ( \th)$ and its royalty cap $ \pi^\ast (\th) \phi$. A lower royalty cap requires a higher upfront price. In the additive noise specification, this menu takes an even simpler form. 

\begin{cor}[Binary menu] \label{BINARYCOR}
Under \cref{AA}, suppose that $N=1$ and that the income distribution follows the additive noise specification. The optimal mechanism $(q^\ast, t^\ast, r^\ast, a^\ast, p^\ast)$ from \cref{MAINTHEOREM} can be implemented by offering the agent a menu with at most two contracts: (i) lump-sum with no royalties or auditing; and possibly (ii) linear royalties at rate $\phi$, certain auditing, and penalties $(\pi - \pi')\phi$.
\end{cor}

The proof gives an explicit formula for the upfront price of each contract. We also provide a condition that characterizes whether the menu has both items. If it does, then high types choose the lump-sum contract, middle types choose the linear royalty, and low types may not buy at all. In the proof, we give a formula for the cutoff types separating these intervals. 
 

\subsection{Comparison with other contingent-payment auctions} \label{sec:comparison}

In our optimal mechanism, the winner's equilibrium payment, as a function of his realized income, is linearly increasing and then flat; see \cref{fig:roy}.  By contrast, in the optimal mechanisms in \cite{demarzo2005bidding} and \cite{inostroza2022screening}, the payment received by the principal, as a function of the realized income, is flat and then linearly increasing. To understand the reason for this difference, it is helpful to consider the obstacle to full extraction in each model. Roughly, the main obstacle in our model is the cost of verification. In each of their models, the obstacle is a form of limited liability, which binds following low income realizations.

Consider the special case of our model in which $\phi_i =1$ for all $i$. In this case, there exists a mechanism that allocates the asset efficiently and leaves no information rents to the agents, but this mechanism requires the principal to always audit the winner. To reduce auditing costs, the optimal mechanism features royalty caps. If the winner claims to owe the full royalty cap, the principal does not audit.

In \cite{demarzo2005bidding}, the project requires an upfront investment by the winner. If this required investment were zero, then full extraction would be possible. With a positive required investment, however, the net cash flow of the project is negative following low income realizations. Thus, full extraction would violate the seller's limited liability constraint. 

In \cite{inostroza2022screening}, there is a single agent who is interpreted as a liquidity supplier for the principal.\footnote{\cite{inostroza2022screening} and \cite{demarzo2005bidding} use opposite sign conventions. In \cite{demarzo2005bidding}, the security specifies what is paid by the bidder to the auctioneer. With this convention, the optimal security is a call option. In \cite{inostroza2022screening}, the principal is selling a security to the agent. Thus, the security specifies what is paid by the principal to the agent. With this convention, the optimal security is a debt contract.} The principal is less patient than the agent. Therefore, the principal would like to receive a larger upfront transfer in exchange for paying out more to the agent in the second period. But the limited liability constraint requires these second-period payments to be smaller than the realized cash flow. 

Our analysis focuses on costly verification rather than liquidity constraints. In our motivating examples, the asset is often transferred because of technological constraints, not liquidity constraints. For example, the government is not liquidity constrained, but it cannot efficiently drill for oil or manage a gambling operation.\footnote{ \citet[Chapter 1]{Contreras2022} lists many reasons other than liquidity constraints for licensing intellectual property.} In our solution, the winner of the asset may face an unlucky income realization that does not cover the upfront transfer that he pays for the asset. 


\subsection{Comparative statics}
   
In this section, we show how the optimal auction in \cref{MAINTHEOREM} depends on the model primitives. Given distribution functions $H_1$ and $H_2$ with respective densities $h_1$ and $h_2$, we say that $H_1$ is weakly larger than $H_2$ with respect to the hazard rate order if 
\[
    \frac{h_1(x)}{1 - H_1(x)} \leq \frac{h_2(x)}{1 - H_2(x)},
\]
for all $x \in \R$.

\begin{thm}[Comparative statics] \label{VV:COMP} For each type $\th_i$, the virtual value $\psi_i (\th_i)$ and the auditing threshold $\pi_i^\ast (\th_i)$ satisfy the following. 
    \begin{enumerate}[label = (\alph*)]
    \item $\psi_i (\th_i)$ is weakly decreasing in $F_i$ with respect to the hazard-rate order.
    \item $\pi_i^\ast (\th_i)$ is weakly increasing in $F_i$ with respect to the hazard-rate order.
    \item $\psi_i (\th_i)$ is weakly decreasing in $c_i$ and weakly increasing in $\phi_i$.
    \item \label{VV:COMP:LIM} $\pi_i^\ast (\th_i)$ is weakly decreasing in $c_i /\phi_i$, provided that $\phi_i > 0$.
    \end{enumerate}
\end{thm}

Suppose that agent $i$'s type distribution $F_i$ increases in the hazard-rate order (and the conditional income distribution for each type remains fixed). Then each fixed type of agent $i$ wins the asset less often (i.e., for a smaller set of type reports of his opponents). And conditional on winning the asset, each type faces a higher royalty cap. The higher royalty cap reduces the information rents of all higher types, who become relatively more common when the type distribution increases with respect to the hazard-rate order.

If the cost $c_i$ of auditing agent $i$ decreases, or the maximal penalty sensitivity $\phi_i$ increases, the asset is allocated to agent $i$ for a larger set of type reports. Moreover, when agent $i$ wins the asset, he faces a higher royalty cap and hence is audited more. In particular, if $\phi_i$ and $c_i$ are scaled up by the same factor (so $c_i/\phi_i$ remains constant), the principal's optimal payoff increases: agent $i$ wins more often, but, conditional on winning, faces the same auditing rule.

Recall the optimal cash auction from \cite{myerson1981optimal}. If the bidders have different valuation distributions, then the winner of the asset is not necessarily the bidder with the highest valuation. The auction favors bidders whose valuation distributions are smaller (in the hazard-rate order).  In our model, there are additional reasons why the asset may be allocated to an agent whose expected income from the asset is not the highest. By \cref{VV:COMP}, the allocation rule favors agents who are cheaper to audit or who can 
be charged steeper penalties (perhaps because their income is more difficult to manipulate). 

\cref{VV:COMP} can also be used to compare the optimal royalty structures in  different environments. Suppose that the asset is a patent license. We expect that the cost of auditing bidders will be higher if the bidders are in a different country than the seller. In this case, our model predicts that the principal will impose lower royalty caps in order to save on auditing costs. 

\section{Conclusion}

We study the optimal design of an auction for an income-generating asset. Our model offers an explanation for the common practice of charging royalties up to a cap. If the winner of the asset claims to owe the full royalty cap, the principal does not audit him. For these income realizations, the cost of auditing outweighs the gains from reducing information rent. 

In our setting, moral hazard is an important consideration, but it is modeled in reduced form, following much of the literature on security design. Incorporating a richer model of moral hazard into the contingent-payment auction setting is an interesting challenge for future work. 


\newpage
\appendix
\section{Proofs} \label{Appendix:proofs}

\subsection{Proof of Lemma \ref{OBJECTIVELEMMA}} \label{PROOFOBJECTIVELEMMA}

Fix a direct mechanism $(q,t,r,a,p)$ that satisfies Condition~\ref{L:punishment_constraint} and the constraints \eqref{IC2},  \eqref{IC1}, and \eqref{IR} for each agent $i$. We can modify the penalty function profile $p$ in such a way that (a) Condition~\ref{L:punishment_constraint} and the constraints \eqref{IC2}, \eqref{IC1}, and \eqref{IR} still hold, and (b) the principal's payoff is unchanged. For each agent $i$ and each history $(\th, \pi_i')$ in $\HH_i$, let
    \[
        \hat{p}_i ( \th, \pi_i', \pi_i) = p_i ( \th, \pi_i', \pi_i') +  (\pi_i - \pi_i')_+ \phi_i.
    \]
Here and below, we use the notation $x_+ = \max \{ x, 0\}$ for any real $x$. By construction, $\hat{p}$ satisfies Condition~\ref{L:punishment_constraint}. For each agent $i$, the modified penalty function $\hat{p}_i$ agrees with $p_i$ whenever $\pi_i = \pi_i'$ and is weakly larger than $p_i$ otherwise. Thus, (a) and (b) hold. Therefore, it suffices to prove \cref{OBJECTIVELEMMA} with $\hat{p}$ in place of $p$. For the rest of the proof, we assume that $p$ has already been replaced with $\hat{p}$. This replacement ensures that $p$ satisfies additional regularity properties. Let $p_{i,3+}$ denote the partial right-derivative of $p_i$ with respect to realized income. For each $(\th, \pi_i)$ in $\HH_i$, we have $p_{i,3+} ( \th, \pi_i, \pi_i) = \phi_i$. We will use this fact in the proof below. 
    
\paragraph{Income-reporting envelope theorem} 

For each agent $i$, define the support
    \[
        S_i = \{ (\th_i, \pi_i) \in \Th_i \times \Pi_i: \pi_i \in \Pi_i (\th_i) \}.
    \]
In the definition of $u_i$ in \eqref{eq:def_ui}, the expression inside the expectation is weakly increasing and $1$-Lipschitz in the true income $\pi_i$, by Condition~\ref{L:punishment_constraint}. We can right-differentiate under the expectation to get
    \begin{equation} \label{eq:D3ui}
         u_{i,3+} (\th_i, \pi_i | \pi_i) = \E_{ \th_{-i}} \Brac{ q_i (\th_i, \th_{-i}) ( 1 - a_i ( \th_i, \th_{-i}, \pi_i) \phi_i ) },
    \end{equation}
for each $(\th_i, \pi_i)$ in $S_i$.
    
By \eqref{IC2}, for each $(\th_i, \pi_i)$ in $S_i$, we have
    \[
      u_i ( \th_i, \pi_i | \pi_i) 
        = \max_{\pi_i' \in \Pi_i (\th_i)} u_i ( \th_i, \pi_i' | \pi_i).
    \]
By \citet[Theorems 1]{milgrom2002envelope}, it can be shown that for each type $\th_i$, there exists a bounded measurable function $m_i ( \th_i, \cdot)$ on $\Pi_i (\th_i)$ satisfying $m_i (\th_i, \pi_i) \geq u_{i,3+} ( \th_i, \pi_i | \pi_i)$ for each $\pi_i$ in $\Pi_i (\th_i)$, with equality whenever the derivative $u_{i,3} ( \th_i, \pi_i | \pi_i)$ exists, such that
    \begin{equation} \label{eq:m_formula}
        u_i ( \th_i, \pi_i | \pi_i)  = u_i ( \th_i, \ubar{\pi}_i (\th_i) | \ubar{\pi}_i (\th_i)) + \int_{ \ubar{\pi}_i ( \th_i)}^{\pi_i}  m_i ( \th_i, z_i) \de z_i,
    \end{equation}
for each $\pi_i$ in $\Pi_i (\th_i)$.\footnote{For each fixed type $\th_i$, the map $\pi_i \mapsto u_i (\th_i, \pi_i | \pi_i)$ is $1$-Lipschitz, so its derivative exists almost everywhere. Extend this almost-everywhere derivative to a measurable $[0,1]$-valued function $\tilde{m}_i (\th_i, \cdot)$ on $\Pi_i (\th_i)$. At each differentiability point $\pi_i$, \citet[Theorem 1]{milgrom2002envelope} implies that $\tilde{m}_i (\th_i, \pi_i) \geq u_{i,3+} ( \th_i, \pi_i | \pi_i)$, with equality if $u_{i,3} ( \th_i, \pi_i | \pi_i)$ exists. Finally, for each $\pi_i \in \Pi_i (\th_i)$, let $m_i(\th_i, \pi_i) = \max\{ \tilde{m}_i (\th_i,\pi_i), u_{i,3+} ( \th_i, \pi_i | \pi_i)\}$.} Extend $m_i$ to $\Th_i \times \Pi_i$ by setting $m_i (\th_i, \pi_i) = Q_i (\th_i)$ if $\pi_i < \ubar{\pi}_i ( \th_i)$ and $m_i ( \th_i, \pi_i) = u_{i,3+} (\th_i, \bar{\pi}_i ( \th_i)| \pi_i)$ if $\pi_i > \bar{\pi}_i ( \th_i)$. Let $\hat{\pi}_i = \inf \Pi_i$. By the definition of $p_i$, it can be checked by cases that
    \begin{equation} \label{eq:m_extended}
          u_i ( \th_i, \proj_{\Pi_i(\th_i)} \pi_i | \pi_i)  = u_i ( \th_i, \ubar{\pi}_i (\th_i) | \hat{\pi}_i) + \int_{\hat{\pi}_i}^{\pi_i}  m_i ( \th_i, z_i) \de z_i,
    \end{equation}
for each $(\th_i, \pi_i)$ in $\Th_i \times \Pi_i$.


\paragraph{Type-reporting envelope theorem}
Recall the definition of $U_i$ from \eqref{eq:Ui}. For any types $\th_i, \th_i' \in \Th_i$, applying \eqref{eq:m_extended} gives
    \begin{equation*}
    \begin{aligned}
        U_i (\th_i' | \th_i) 
        &= \int_{\hat{\pi}_i}^{\infty} u_i (\th_i', \proj_{\Pi_i (\th_i')} \pi_i| \pi_i) g_i( \pi_i | \th_i) \de \pi_i \\
        &= u_i ( \th_i', \ubar{\pi}_i (\th_i') | \hat{\pi}_i) + \int_{ \hat{\pi}_i}^{\infty} \Paren{\int_{\hat{\pi}_i}^{\pi_i} m_i (\th_i', z_i) \de z_i} g_i( \pi_i |\th_i) \de \pi_i.
    \end{aligned}
    \end{equation*}
Change the order of integration (and switch the variable labels) to get\footnote{This integral is well-defined since $m_i$ is bounded.}
    \begin{equation*} 
        U_i (\th_i' | \th_i) =u_i ( \th_i', \ubar{\pi}_i (\th_i') | \hat{\pi}_i) + \int_{ \hat{\pi}_i}^{\infty} (1 - G_i (\pi_i | \th_i))  m_i (\th_i', \pi_i) \de \pi_i.
    \end{equation*}
Differentiate under the integral with respect to $\th_i$ to get\footnote{We can differentiate under the integral because $m_i$ is bounded and the functions $G_i  (\pi_i | \cdot)$ are uniformly Lipschitz, by assumption.} 
    \[
        U_{i,2} (\th_i| \th_i) = \int_{ \hat{\pi}_i}^{\infty} -G_{i,2}  (\pi_i | \th_i) m_i (\th_i, \pi_i) \de \pi_i.
    \]
    
By \eqref{IC1}, for each $\th_i$ in $\Th_i$, we have
    \[
        U_i ( \th_i| \th_i) - T_i (\th_i)  
        = \max_{\th_i' \in \Th_i} \Set{ U_i ( \th_i' | \th_i) - T_i (\th_i')}.
    \]
By our Lipschitz condition on $G_i ( \pi_i| \cdot)$, we can apply the envelope theorem \citep{milgrom2002envelope}. For all $\th_i$ in $\Th_i$, we have
    \begin{equation} \label{eq:Lem_equality}
    \begin{aligned}
       &  U_i ( \th_i| \th_i) - T_i (\th_i)  \\
        &= U_i (\ubar{\th}_i | \ubar{\th}_i) - T_i ( \ubar{\th}_i) + \int_{\ubar{\th}_i}^{\th_i} \Paren{ \int_{\hat{\pi}_i}^{\infty} -G_{i,2} (\pi_i| y_i) m_i (y_i, \pi_i) \de \pi_i } \de y_i \\
        &\geq \int_{\ubar{\th}_i}^{\th_i} \Paren{ \int_{\ubar{\pi}_i (y_i)}^{\bar{\pi}_i(y_i)} -G_{i,2} (\pi_i| y_i)  u_{i,3+} (y_i, \pi_i| \pi_i) \de \pi_i } \de y_i,
    \end{aligned} 
    \end{equation}
where we get the final inequality by applying \eqref{IR} and noting that $G_{i,2}$ vanishes outside $S_i$ and that $m_i ( y_i, \pi_i) \geq u_{i,3+} ( y_i, \pi_i |\pi_i)$ for $(y_i, \pi_i)$ in $S_i$.\footnote{Note that the function $m_i$ is guaranteed to be measurable in its second argument only, but this is sufficient for the proof.}

Now take an expectation over $\th_i$, change the order of integration, and multiply and divide by $f_i (\th_i) g_i (\pi_i |\th_i)$ to conclude that
    \begin{equation} \label{eq:representation}
    \begin{aligned}
        \E  [ U_i ( \th_i| \th_i) - T_i (\th_i)] 
        &\geq \E \Brac{  \mu_i (\th_i, \pi_i)  u_{i,3+} (\th_i, \pi_i| \pi_i)} \\
        &= \E \Brac{ q_i (\th_i, \th_{-i}) \mu_i (\th_i, \pi_i) ( 1 - a_i ( \th_i, \th_{-i}, \pi_i) \phi_i ) },
    \end{aligned}
    \end{equation}
where we have used the independence of $(\th_i, \pi_i)$ and $\th_{-i}$.

\paragraph{Principal's payoff} We can express the principal's expected payoff in terms of the functions $U_i$ and $T_i$ as
    \[
      \E_{\th}  \Brac{ \sum_{i=1}^{N}  \Paren{ q_i (\th) \E_{\pi_i | \th_i} \Brac{ \pi_i  - c_i a_i (\th, \pi_i)} -  U_i (\th_i | \th_i) + T_i (\th_i)}}.
    \]
Plug in \eqref{eq:representation} and simplify to get the upper bound
    \begin{equation}
    \label{eq:unnormalized}
      \E_{\th}  \Brac{ \sum_{i=1}^{N}   q_i (\th) \E_{\pi_i | \th_i} \Brac{ \pi_i  -\mu_i (\th_i, \pi_i)  + a_i (\th, \pi_i) ( \mu_i (\th_i, \pi_i) \phi_i - c_i) }}.
    \end{equation}
To get the desired expression, it remains to check that
    \begin{equation} \label{eq:Myerson_identity}
        \E_{\pi_i | \th_i} [ \pi_i - \mu_i (\th_i, \pi_i)] = \th_i - \frac{1 - F_i (\th_i)}{f_i (\th_i)}.
    \end{equation}
By our normalization and a standard probability identity,
   \[
   \th_i = \E[ \pi_i |  \th_i] = \int_{0}^{\infty} (1 - G_i (\pi_i |  \th_i)) \de \pi_i.
   \]
Differentiating under the integral sign gives
    \begin{equation} \label{eq:identity}
       1 =  -\int_{0}^{\infty} G_{i,2} (\pi_i| \th_i) \de \pi_i = 
       \E_{\pi_i | \th_i} \Brac { \frac{-G_{i,2} (\pi_i| \th_i)}{g_i (\pi_i | \th_i)}},
    \end{equation}
as needed.


\subsection{Regularity assumptions} \label{A:SHIFT}

In our notation, the regularity assumptions in \citet[p.~709]{esHo2007optimal} state that for each agent $i$,
\begin{enumerate}
    \item $f_i (\th_i)/ (1 - F_i(\th_i))$ is weakly increasing in $\th_i$;
    \item $G_{i,2} (\pi_i | \th_i) / g_i (\pi_i | \th_i)$ is weakly increasing in $\th_i$ and $\pi_i$.
\end{enumerate}
We prove that these regularity assumptions imply \cref{AA}.\footnote{In fact, the regularity assumptions in \cite{esHo2007optimal}, together with the type normalization \eqref{eq:normalized}, imply additive full-support noise; see \cite{BP2024note}.} Recall that 
\[
    \mu_i (\th_i, \pi_i) = - \frac{G_{i,2} ( \pi_i | \th_i)}{g_i(\pi_i | \th_i)} \cdot \frac{1 - F_i (\th_i)}{f_i(\th_i)}.
\]
Thus, $\mu_i ( \th_i, \pi_i)$ is weakly decreasing in $\th_i$ and $\pi_i$. \cref{AA}.\ref{it:single-crossing} follows immediately. For \cref{AA}.\ref{it:regularity}, substitute in \eqref{eq:Myerson_identity} to see that
\[
    \psi_i (\th_i) 
    = \th_i - \E_{\pi_i| \th_i} \Brac{ \mu_i (\th_i, \pi_i) - ( \mu_i(\th_i, \pi_i) \phi_i - c_i)_+}.
\]
The expression inside the expectation is decreasing in $\th_i$ and $\pi_i$, and we have $G_{i,2} ( \pi_i | \th_i) < 0$. Thus, the expectation is decreasing in $\th_i$, and hence, $\psi_i (\th_i)$ is strictly increasing in $\th_i$. 

Next, we provide an example of distributions $G_i$ and $F_i$ for which \cref{AA} holds, but the ratio
\[
\frac{G_{i,2}(\pi_i \mid \th_i)}{g_i(\pi_i \mid \th_i)}
\]
is strictly \emph{decreasing} in $\th_i$, violating the regularity assumptions in \cite{esHo2007optimal}. We consider a single agent, and we drop agent indices. Assume that $\th$ is uniformly distributed on $[1,2]$. Given $\th$, the conditional distribution of $\pi$ is uniform over $[\th - e^{-\th/2}, \th + e^{-\th/2}]$. For each $\th \in [1,2]$ and $\pi \in (\th - e^{-\th/2}, \th + e^{-\th/2})$, it can be verified that
\[
\frac{G_2(\pi | \th)}{g(\pi| \th)}
=
\frac{\pi-\th}{2}-1,
\]
which is strictly increasing in $\pi$ and strictly \textit{decreasing} in $\th$. However, we have
\[
\mu(\th, \pi)
=(2-\th)\left(1-\frac{\pi-\th}{2}\right),
\]
which is weakly decreasing in both arguments. This implies that the virtual value $\psi$ is strictly increasing. Hence, \cref{AA} is satisfied.


\subsection{Transformed additive noise} \label{sec:proof_transformed_additive}

Let $H_i$ and $h_i$ denote the cumulative distribution function and probability density function of $\e_i$, respectively. In place of $F_i, f_i, G_i, g_i, \mu_i$, we write $\hat{F}_i, \hat{f}_i, \hat{G}_i, \hat{g}_i, \hat{\mu}_i$ for the corresponding functions defined on unnormalized signals. We have
\[
    \hat{G}_i (\pi_i | \g_i) = \P_{\e_i} \Paren {\a_i ( \g_i + \e_i) \leq \pi_i}   = H_i ( \a_i^{-1} (\pi_i) - \g_i ).
\]
Therefore, 
\begin{equation*}
\begin{aligned}
        \hat{G}_{i,2} ( \pi_i | \g_i) &= - h_i ( \a_i^{-1} (\pi_i) - \g_i) \\ 
    \hat{g}_i (\pi_i | \g_i) &= \frac{h_i ( \a_i^{-1} (\pi_i) - \g_i)}{\a_i' (\a_i^{-1}(\pi_i))}. 
\end{aligned}
\end{equation*}

First, we prove \cref{AA}.\ref{it:single-crossing}. We have
\[
    \hat{\mu}_i (\g_i, \pi_i) = - \frac{  \hat{G}_{i,2} ( \pi_i | \g_i)}{ \hat{g}_i (\pi_i | \g_i)} \cdot \frac{1 - \hat{F}_i(\g_i)}{\hat{f}_i(\g_i)} = \a_i' ( \a_{i}^{-1} (\pi_i)) \cdot \frac{1 - \hat{F}_i(\g_i)}{\hat{f}_i(\g_i)}.
\]
Notice that the right side is multiplicatively separable in $\pi_i$ and $\g_i$. The function $\hat{\mu}_i$ is weakly decreasing in $\pi_i$ since $\a_i' > 0$ and $\a_i'' \leq 0$. And $\hat{\mu}_i$ is weakly decreasing in $\g_i$ because the distribution of $\g_i$ has weakly increasing hazard rate.  It is easily verified that $\hat{\mu}_i (\g_i, \pi_i) = \mu_i( \E[ \pi_i | \g_i], \pi_i)$, so \cref{AA}.\ref{it:single-crossing} follows. 

Next, we prove  \cref{AA}.\ref{it:regularity}. Since we have not normalized the signal,  we see from \eqref{eq:unnormalized} that the associated virtual value (with respect to the unnormalized signal) is given by
\begin{equation*}
\begin{aligned}
    \hat{\psi}_i (\g_i) 
    &=  \E_{\pi_i | \g_i} \Brac{ \pi_i -  \hat{\mu}_i (\g_i, \pi_i) + (\hat{\mu}_i (\g_i, \pi_i) \phi_i - c_i)_+} \\
    &=  \E_{\pi_i | \g_i} \Brac{ \max\{ \pi_i - (1 - \phi_i) \hat{\mu}_i ( \g_i , \pi_i) - c_i, \pi_i -  \hat{\mu}_i ( \g_i , \pi_i)\} }.
\end{aligned}
\end{equation*}
We claim that $\hat{\psi}_i$ is strictly increasing in $\g_i$. Inside the expectation, each function inside the maximum is strictly increasing in $\pi_i$ and weakly increasing in $\g_i$ (since the distribution of $\g_i$ has weakly increasing hazard rate). Moreover, the conditional distribution of $\pi_i$ given $\g_i$ is strictly increasing in $\g_i$,  with respect to first-order stochastic dominance. Thus, $\hat{\psi}_i$ is strictly increasing. It is easily verified that $\hat{\psi}_i (\g_i) = \psi_i ( \E[\pi_i | \g_i])$, so \cref{AA}.\ref{it:regularity} follows. 


\subsection{Proof of Theorem \ref{MAINTHEOREM}} \label{PROOFMAINTHEOREM}

Before the main proof, we first confirm that $\pi_i^\ast$ is weakly decreasing.
  Recall that $\pi_i^\ast$ is defined as follows. Here, we use the notation $\D_i (\th_i, \pi_i) = \mu_i (\th_i, \pi_i) \phi_i - c_i$. For each $\th_i \in \Th_i$, let
  \[
    \Pi_i^\ast(\th_i) = \{ \pi_i \in (\ubar{\pi}_i (\th_i), \bar{\pi}_i (\th_i)) : \D_i (\th_i, \pi_i) \geq 0 \}.
\]
Define $\pi_i^\ast \colon \Th_i \to [0,\infty]$ by 
\begin{equation*}
    \pi_i^\ast (\th_i)
    = 
    \begin{cases}
        \infty &\text{if}~\Pi_i^\ast (\th_i) = (\ubar{\pi}_i (\th_i), \bar{\pi}_i (\th_i)),\\
     \sup \Pi_i^\ast( \th_i)  &\text{if}~     \varnothing \subsetneq   \Pi_i^\ast (\th_i) \subsetneq (\ubar{\pi}_i (\th_i), \bar{\pi}_i (\th_i)), \\
       0 &\text{if}~\Pi_i^\ast (\th_i) = \varnothing.
    \end{cases}
\end{equation*}

To prove that $\pi_i^\ast$ is weakly decreasing, it suffices to show the following local monotonicity property: for any types $\th_i$ and $\th_i'$ with $\th_i < \th_i'$ and $\bar{\pi}_i (\th_i) > \ubar{\pi}_i (\th_i')$, we have $\pi_i^\ast(\th_i) \geq \pi_i^\ast (\th_i')$. It then follows from a compactness argument that $\pi_i^\ast$ is globally weakly decreasing.\footnote{Since $\ubar{\pi}_i$ and $\bar{\pi}_i$ are continuous, for each type $\th_i$ there is an open interval $U_{\th_i}$ containing $\th_i$ such that for all types $\th_i', \th_i'' \in U_{\th_i}$ with $\th_i' < \th_i''$ we have $\bar{\pi}_i (\th_i') > \ubar{\pi}_i (\th_i'')$. The local monotonicity property implies that $\pi_i^\ast$ is weakly decreasing over $U_{\th_i}$. The collection $\{ U_{\th_i} \}_{\th_i \in \Th_i}$ is an open cover of $\Th_i$. Since $\Th_i$ is compact, $\{ U_{\th_i} \}_{\th_i \in \Th_i}$ has a finite subcover. Global monotonicity follows because any two types can be connected by a finite path where each pair of consecutive types lies in some interval in the finite cover.} Fix types $\th_i$ and $\th_i'$ with $\th_i < \th_i'$ and $\bar{\pi}_i (\th_i) > \ubar{\pi}_i (\th_i')$. Suppose for a contradiction that $\pi_i^\ast(\th_i) < \pi_i^\ast (\th_i')$. In particular, $\pi_i^\ast (\th_i) < \infty$ and  $\pi_i^\ast (\th_i')>0$, so the intervals $(\pi_i^\ast(\th_i), \bar{\pi}_i (\th_i))$ and $(\ubar{\pi}_i (\th_i'), \pi_i^\ast (\th_i'))$ are nonempty. Moreover, these intervals have nonempty intersection because $\bar{\pi}_i (\th_i) > \ubar{\pi}_i (\th_i')$ and $\pi_i^\ast (\th_i) < \pi_i^\ast (\th_i')$. Choose $\pi_i \in (\pi_i^\ast(\th_i), \bar{\pi}_i (\th_i)) \cap (\ubar{\pi}_i (\th_i'), \pi_i^\ast (\th_i'))$.  By the definition of $\pi_i^\ast (\th_i')$, there exists $\pi_i'$ in $(\pi_i,  \pi_i^\ast (\th_i') \wedge \bar{\pi}_i (\th_i'))$ with $\D_i (\th_i', \pi_i') \geq 0$. Since $\D_i$ is single-crossing from above in income, we have $\D_i (\th_i', \pi_i) \geq 0$. But we also have $\D_i ( \th_i, \pi_i) < 0$ because $\pi_i$ is in $(\pi_i^\ast(\th_i), \bar{\pi}_i (\th_i))$. This contradicts the assumption that $\D_i$ is single-crossing from above in the type. 

Now we turn to the main proof. For any mechanism $(q,t,r,a,p)$ satisfying Condition~\ref{L:punishment_constraint} and the constraints \eqref{IC2}, \eqref{IC1}, and \eqref{IR} for each agent $i$, the principal's expected payoff, denoted $V ( q,t,r,a,p)$, satisfies 
    \begin{align} 
        V ( q,t,r,a,p) 
        &\leq \E \Brac{ \sum_{i=1}^{N} q_i (\th) \Psi_i (\th)} \label{ineq_1} \\
         &\leq \E \Brac{ \max_{i} \Psi_i (\th)_+} \label{ineq_2} \\
        &\leq \E \Brac{ \max_{i} \psi_i (\th_i)_+}, \label{ineq_3}
    \end{align}
where \eqref{ineq_1} follows from \cref{OBJECTIVELEMMA} and \eqref{ineq_3} follows from the definitions of $\Psi_i$ and $\psi_i$.\footnote{For any function $h$, we denote $(h(x))_+$ by $h(x)_+$.} Note that $\Psi_i$ depends on the auditing rule $a_i$.
     
For the mechanism $( q,t,r,a,p) = (q^\ast, t^\ast, r^\ast, a^\ast, p^\ast)$, we claim that the inequalities \eqref{ineq_1}--\eqref{ineq_3} hold with equality. For \eqref{ineq_1}, follow the proof of \cref{OBJECTIVELEMMA} with  $( q,t,r,a,p) = (q^\ast, t^\ast, r^\ast, a^\ast, p^\ast)$.\footnote{Except in this case, do not replace $p^\ast$ with the modified penalty rule $\hat{p}$.}  The inequality in  \eqref{eq:Lem_equality} holds with equality because (a) \eqref{IR} holds with equality for the lowest type of each agent, and (b) for each $(\th_i, \pi_i) \in S_i$ with $\pi_i \neq \pi_i^\ast (\th_i)$, the partial derivative $u_{i,3} ( \th_i, \pi_i | \pi_i)$ exists and agrees with the expression in \eqref{eq:D3ui}. For \eqref{ineq_2}, equality is immediate from the definition of $q^\ast$. For \eqref{ineq_3}, since $\mu_i ( \th_i, \pi_i) \phi_i - c_i$ is single-crossing from above in $\pi_i$ (by \cref{AA}.\ref{it:single-crossing}), note that
    \[
       [\mu_i ( \th_i, \pi_i) \phi_i - c_i > 0]  \leq a_i^\ast (\th, \pi_i) \leq [\mu_i ( \th_i, \pi_i) \phi_i - c_i \geq 0].
    \]
To complete the proof, we show that $(q^\ast, t^\ast, r^\ast, a^\ast, p^\ast)$ is dominant-strategy incentive compatible and dominant-strategy individually rational. For the rest of the proof, we focus on a fixed agent $i$.

First, we prove dominant-strategy individual rationality.  Given a report vector $\th_{-i}$ in $\Th_{-i}$ from agent $i$'s opponents, the expected utility for type $\th_i$ from truthfully reporting his type and subsequent income is 
\[
 \int_{\ul{\th}_i}^{\th_i} q_i^\ast(z_i,\th_{-i})(1-\Phi_i(z_i))  \de z_i.
\]
This expression is nonnegative because $\Phi_i(z_i) \leq 1$ for all $z_i$. 

Next, we prove dominant-strategy incentive compatibility. If agent $i$ wins the asset, the royalty and the penalty do not depend on the other agents' reports. We check that at the income-reporting stage, it is optimal for agent $i$ to report his income as truthfully as possible after any type report $\th_i'$. To see this, suppose agent $i$'s true income is $\pi_i$. Consider agent $i$'s additional (post-transfer) payment as a function of his report $\pi_i'$. If  $\pi_i' < \pi_i^\ast (\th_i')$, then he pays $\min\{ \pi_i, \pi_i^\ast(\th_i') \} \phi_i$. If  $\pi_i' \geq \pi_i^\ast (\th_i')$, then he pays $\pi_i^\ast(\th_i') \phi_i$. Therefore, reporting $\pi_i' = \proj_{\Pi_i(\th_i')} \pi_i$ is optimal because $\proj_{\Pi_i(\th_i')} \pi_i \geq \pi_i^\ast (\th_i')$ if and only if $\pi_i \geq \pi_i^\ast (\th_i')$.

Now we consider the type-reporting stage. Fix a report vector $\th_{-i}$ in $\Th_{-i}$ from agent $i$'s opponents. 
Given the report $\th_{-i}$, the difference in expected utility for type $\th_i$ between reporting type $\th_i$ (and then reporting income truthfully) and reporting type $\th_i'$ (and then reporting income as truthfully as possible) can be expressed as
    \begin{multline} \label{eq:nonnegative}
    \int_{\th_i'}^{\th_i} q_i^\ast (z_i, \th_{-i}) [ 1 - \Phi_i ( z_i)] \de z_i  \\ 
        + q_i^\ast (\th_i', \th_{-i}) \int_{0}^{\infty} \bigl( \pi_i - \phi_i \min \{ \pi_i, \pi_i^\ast ( \th_i')\} \bigr) \bigl[ g_i(\pi_i | \th_i') - g_i(\pi_i | \th_i) \bigr] \de \pi_i.
    \end{multline}
Note that the sign of the first integral depends on the ordering of $\th_i$ and $\th_i'$. For any $z_i$ in $\Th_i$ and $\pi_i$ in $\Pi_i$, let
    \[
        m_i^\ast ( z_i, \pi_i ;\th_{-i}) = q_i^\ast (z_i, \th_{-i}) \Paren{ 1 - \phi_i [ \pi_i \leq \pi_i^\ast (z_i)]}.
    \]
Since $q_i^\ast (z_i, \th_{-i})$ is weakly increasing in $z_i$, and $\pi_i^\ast (z_i)$ is weakly decreasing in $z_i$, it follows that $m_i^\ast$ is weakly increasing in $z_i$. We will express each line of \eqref{eq:nonnegative} in terms of $m_i^\ast$. 

From the definition of $\Phi_i$ in \eqref{eq:Psi}, and the implication \eqref{eq:identity} of our normalization, we have
    \[
        1 - \Phi_ i (z_i) = \int_{0}^{\infty} -G_{i,2} (\pi_i | z_i) \Paren{1 - \phi_i [ \pi_i \leq \pi_i^\ast (z_i)]}  \de \pi_i,
    \]
so the first line in \eqref{eq:nonnegative} can be expressed as
    \begin{equation} \label{eq:first}
     \int_{\th_i'}^{\th_i} \int_{0}^{\infty}   -G_{i,2} ( \pi_i | z_i)   m_i^\ast (z_i, \pi_i; \th_{-i}) \de \pi_i  \de z_i.
    \end{equation}
For the integral in the second line of \eqref{eq:nonnegative}, integrating by parts gives\footnote{Integration by parts still holds when the ``antiderivative'' is an absolutely continuous function with almost-sure derivative.}
    \[
        \int_{0}^{\infty} \Paren{ 1- \phi_i [ \pi_i \leq \pi_i^\ast (\th_i')]} \Paren{ G_i ( \pi_i | \th_i) - G_i ( \pi_i | \th_i')} \de \pi_i,
    \]
so the second line of \eqref{eq:nonnegative} becomes
    \begin{equation} \label{eq:second}
    \int_{\th_i'}^{\th_i} \int_{0}^{\infty}   G_{i,2} ( \pi_i | z_i)   m_i^\ast (\th_i', \pi_i; \th_{-i}) \de \pi_i  \de z_i.
    \end{equation}
Putting together  \eqref{eq:first} and \eqref{eq:second}, the expression in \eqref{eq:nonnegative} becomes
    \[
         \int_{\th_i'}^{\th_i} \int_{0}^{\infty}   -G_{i,2} ( \pi_i | z_i)  \Paren{  m_i^\ast (z_i, \pi_i; \th_{-i}) -   m_i^\ast (\th_i', \pi_i; \th_{-i})} \de \pi_i  \de z_i.
    \]
This expression is nonnegative since $-G_{i,2}$ is nonnegative and $m_i^\ast$ is weakly increasing in its first argument. 

\paragraph{Crossing payments} Here, we prove the crossing property referenced after      \cref{fig:roy}. First, we introduce some notation. For all $\th \in \Th$ with $q_i^\ast (\th) = 1$, let 
    \begin{equation*}
       s_i^\ast (\th, \pi_i) = t_i^\ast ( \th) + r_i^\ast (\th, \proj_{\Pi_i(\th_i)} \pi_i ) + a_i^\ast (\th,\proj_{\Pi_i(\th_i)} \pi_i) p_i^\ast ( \th, \proj_{\Pi_i(\th_i)} \pi_i, \pi_i).
    \end{equation*}
Thus, $s_i^\ast ( \th, \pi_i)$ is agent $i$'s total ex-post payment when the type vector $\th$ is reported, agent $i$'s true income is $\pi_i$, and agent $i$ reports income $\proj_{\Pi_i(\th_i)} \pi_i$. Observe that $\proj_{\Pi_i(\th_i)} \pi_i \geq \pi_i^\ast ( \th_i)$ if and only if $\pi_i \geq \pi_i^\ast (\th_i)$. Therefore, for each $\th \in \Th$ with $q_i^\ast (\th) = 1$ and each $\pi_i \in \Pi_i$, we have
\[
       s_i^\ast (\th, \pi_i)    = t_i^\ast(\th) + \min \{ \pi_i, \pi_i^\ast(\th_i)\} \phi_i.
    \]
Note that \cref{fig:roy} plots the function $s_i^\ast(\th_i', \ubar{\th}_{-i}, \cdot)$ over the domain $\Pi_i (\th_i')$ and the function $s_i^\ast(\th_i'', \ubar{\th}_{-i}, \cdot)$ over the domain $\Pi_i (\th_i'')$. 

Consider agent $i$. Fix $\th_{-i} \in \Th_{-i}$ and $\th_i', \th_i'' \in \Th_i$ with $\th_i' < \th_i''$. Suppose that $q_i^\ast ( \th_i', \th_{-i}) = 1$ and $q_i^\ast ( \th_i'', \th_{-i}) = 1$. Since $\pi_i^\ast$ is weakly decreasing, we have $\pi_i^\ast (\th_i') \geq \pi_i^\ast (\th_i'')$. We compare the functions $s_i^\ast ( \th_i', \th_{-i}, \cdot)$ and $s_i^\ast ( \th_i'', \th_{-i}, \cdot)$ over $\Pi_i$. We claim that if $\pi_i^\ast ( \th_i') $ and $\pi_i^\ast (\th_i'')$ are both in $(0, \infty)$, then either (i) $\pi_i^\ast (\th_i') = \pi_i^\ast (\th_i'')$, in which case the functions agree everywhere, or (ii) $\pi_i^\ast (\th_i') > \pi_i^\ast (\th_i'')$, in which case the functions intersect at exactly one point, and this point is in $(\pi_i^\ast (\th_i''), \pi_i^\ast (\th_i'))$.

Now we verify the claim. Since $q_i^\ast( \th_i', \th_{-i}) = q_i^\ast ( \th_i'', \th_{-i})$, dominant-strategy incentive compatibility implies that
    \begin{equation} \label{eq:DSIC_s}
    \begin{aligned}
         \E_{\pi_i | \th_i'} [s_i^\ast ( \th_i', \th_{-i}, \pi_i)] &\leq \E_{\pi_i | \th_i'} [s_i^\ast ( \th_i'', \th_{-i}, \pi_i)], \\
        \E_{\pi_i | \th_i''} [s_i^\ast ( \th_i'', \th_{-i}, \pi_i)] &\leq \E_{\pi_i | \th_i''}[ s_i^\ast ( \th_i', \th_{-i}, \pi_i)].
    \end{aligned}
    \end{equation} 
Suppose that $\pi_i^\ast ( \th_i') $ and $\pi_i^\ast (\th_i'')$ are both in $(0, \infty)$. If $\pi_i^\ast (\th_i') =\pi_i^\ast ( \th_i'')$, then the curves must agree everywhere; otherwise, one curve is pointwise strictly larger, contrary to \eqref{eq:DSIC_s}. If $ \pi_i^\ast (\th_i') > \pi_i^\ast ( \th_i'')$, then by comparing slopes, we see that over the interval $[\pi_i^\ast (\th_i''), \pi_i^\ast(\th_i')]$, the curves intersect at exactly one point. We show that this point cannot be either endpoint. If the intersection is at $\pi_i^\ast (\th_i'')$, then $s_i^\ast ( \th_i', \th_{-i}, \cdot)$ is pointwise weakly larger than $s_i^\ast ( \th_i'', \th_{-i}, \cdot)$. By \eqref{eq:DSIC_s}, we must have $\Pi_i (\th_i') \subseteq (-\infty, \pi_i^\ast (\th_i'')]$. In particular,  $\bar{\pi}_i ( \th_i') \leq \pi_i^\ast ( \th_i'') <  \pi_i^\ast ( \th_i')$, so $\pi_i^\ast ( \th_i') = \infty$, which is a contradiction. If the intersection is at $\pi_i^\ast (\th_i')$, then  $s_i^\ast ( \th_i'', \th_{-i}, \cdot)$ is pointwise weakly larger than $s_i^\ast ( \th_i', \th_{-i}, \cdot)$. By \eqref{eq:DSIC_s}, we must have $\Pi_i (\th_i'') \subseteq [\pi_i^\ast (\th_i'), \infty)$. In particular, $\pi_i^\ast (\th_i'') < \pi_i^\ast (\th_i') \leq \ubar{\pi}_i (\th_i'')$, so $\pi_i^\ast ( \th_i'') = 0$, which is a contradiction. 


\subsection{Noisy auditing} \label{sec:noisy_auditing}

Let $Z_i$ denote the support of the signal $\z_i$. With noisy auditing, the principal chooses a  penalty rule $\tilde{p}_i \colon \HH_i \times Z_i \to \R$, for each agent $i$. We require that these penalty rules satisfy the analog of Condition~\ref{L:punishment_constraint}, with $\z_i$ in place of $\pi_i$.

Given a penalty rule $\tilde{p}_i$, define the associated \emph{effective penalty rule} $p_i \colon \HH_i \times \Pi_i \to \R$ by 
\[
    p_i (\th, \pi_i', \pi_i) = \E_{ \z_i | \pi_i} [ \tilde{p}_i (\th, \pi_i', \z_i)].
\]
Using penalty rule $\tilde{p}_i$ with noisy auditing is equivalent (in terms of expected payments and utilities) to using the associated effective penalty rule $p_i$ in the original model with perfect auditing. In the main text, we present a mechanism in the noisy auditing model that achieves the principal's maximum payoff in the perfect auditing model. To prove that this mechanism is optimal, it suffices to check that noisy auditing (weakly) shrinks the principal's feasible set. 

Formally, we show that if $\tilde{p}_i$ satisfies Condition~\ref{L:punishment_constraint}, then the associated effective penalty rule $p_i$ satisfies Condition~\ref{L:punishment_constraint} as well. Let $(\th', \pi_i')$ be a report history in $\HH_i$. Fix $\pi_i, \hat{\pi}_i \in \Pi_i$ with $\hat{\pi}_i > \pi_i$. The associated conditional distributions of $\z_i$ are ordered by first-order stochastic dominance, so we can construct a random vector $(\z_i, \hat{\z}_i)$ satisfying $\z_i \leq \hat{\z}_i$ such that the marginal distribution of $\z_i$ (respectively, $\hat{\z}_i$) agrees with the conditional signal distribution, given $\pi_i$ (respectively $\hat{\pi}_i$). By Condition~\ref{L:punishment_constraint}, we have
\[
    0 \leq \tilde{p}_i ( \th', \pi_i', \hat{\z}_i) - \tilde{p}_i ( \th', \pi_i', \z_i) \leq  (\hat{\z}_i - \z_i) \phi_i.
\]
Taking expectations gives
\[
    0 \leq p_i ( \th', \pi_i', \hat{\pi}_i) - p_i ( \th', \pi_i', \pi_i) \leq (\hat{\pi}_i - \pi_i) \phi_i, 
\]
as desired. 


\subsection{Proof of Corollary \ref{BINARYCOR}} 

We drop the agent subscripts. Under the additive noise specification, 
\[
    \mu (\th, \pi) = \frac{1 - F (\th)}{f(\th)}. 
\]
Define the cutoff types by
\begin{equation*}
\begin{aligned}
    \th^\ast &= \sup \Big\{ \th \in \Th : \frac{1 - F(\th)}{f(\th)} \phi - c \geq 0 \Big\}, \\
    \th_0 &= \inf\Big\{ \th \in \Th : \th - \frac{1 - F(\th)}{f(\th)} + \left( \frac{1 - F(\th)}{f(\th)} \phi - c\right)_+ > 0 \Big\},
\end{aligned}
\end{equation*}
where we follow the convention that $\sup \varnothing = -\infty$.  There are two cases. 

If $\th_0 \geq \th^\ast$, then the mechanism $(q^\ast, t^\ast, r^\ast, a^\ast, p^\ast)$ from \cref{MAINTHEOREM} can be implemented by offering a singleton menu with a lump-sum contract at price $\th_0$. 

If $\th_0 < \th^\ast$, then the mechanism $(q^\ast, t^\ast, r^\ast, a^\ast, p^\ast)$ from \cref{MAINTHEOREM} can be implemented by offering a binary menu with a lump-sum contract at price $(1 - \phi) \th_0 + \phi \th^\ast$ and a linear royalty contract at price $(1 - \phi) \th_0$. 


\subsection{Proof of Theorem \ref{VV:COMP}}

Using the identity \eqref{eq:identity}, we can express agent $i$'s virtual value as 
\begin{equation*} 
\begin{aligned}
    \psi_i (\th_i) 
    &= \th_i - \E_{\pi_i |\th_i} [ \mu_i (\th_i, \pi_i)]  + \E_{\pi_i | \th_i} \Brac{ (\mu_i(\th_i, \pi_i) \phi_i - c_i)_+} \\
    &=  \E_{\pi_i | \th_i} \Brac{ \max\{  \pi_i - (1 - \phi_i) \mu_i ( \th_i , \pi_i) - c_i, \pi_i - \mu_i ( \th_i , \pi_i)\} }.
\end{aligned}
\end{equation*}
Therefore, $\psi_i (\th_i)$ is weakly decreasing in $c_i$ and weakly increasing in $\phi_i$ (since $\mu_i$ is nonnegative). Moreover, $\psi_i (\th_i)$ is 
weakly decreasing in $\mu_i (\th_i, \pi_i)$, which is increasing in the hazard-rate order. Next, the comparative statics for $\pi_i^\ast (\th_i)$ are immediate upon noting that whenever $\phi_i > 0$,  the auditing condition can be expressed as
\begin{equation} \label{eq:auditing_condition}
    \mu_i (\th_i, \pi_i) \geq c_i / \phi_i.
\end{equation}
If $F_i$ increases in the hazard-rate order, then $\mu_i (\th_i, \pi_i)$ weakly increases, so \eqref{eq:auditing_condition} is relaxed and $\pi_i^\ast(\th_i)$ weakly increases. Conversely, if $c_i /\phi_i$ increases, then \eqref{eq:auditing_condition} becomes more restrictive, so $\pi_i^\ast( \th_i)$ weakly decreases. 


\newpage
	\begingroup
	\setlength{\bibsep}{1pt}
        \small
	\bibliographystyle{aer}
        \bibliography{Procurements.bib}
	\endgroup


\end{document}